\documentclass[11pt]{article}
\usepackage[top=1in, bottom=1in, left=1in, right=1in]{geometry}
\usepackage{tikz}
\usetikzlibrary{shapes,arrows,matrix,patterns,positioning,calc,snakes}
\usepackage{color}
\usepackage{verbatim}
\usepackage{graphicx}
\usepackage{algorithmic}
\usepackage{amssymb}
\usepackage{epstopdf}
\usepackage[linesnumbered,boxed,ruled,commentsnumbered]{algorithm2e}
\usepackage{float}
\usepackage{amsmath,amsthm,bm,color,epsfig,enumerate,caption}
\usepackage{hyperref}
\usepackage{booktabs}
\usepackage{multirow}
\usepackage{array}
\usepackage{tabu}
\usepackage{breqn}
\usepackage{mathtools}
\usepackage{multicol}
\usepackage{multirow}
\numberwithin{equation}{section}
 \usepackage{float}

\newtheorem{theorem}{Theorem}[section]

\newtheorem{lemma}[theorem]{Lemma}

\renewcommand{\appendix}[1]{
\section*{Appendix: #1}
}

\makeatletter
\newcommand*{\extendadd}{
  \mathbin{
    \mathpalette\extend@add{}
  }
}
\newcommand*{\extend@add}[2]{
  \ooalign{
    $\m@th#1\leftrightarrow$%
    \vphantom{$\m@th#1\updownarrow$}
    \cr
    \hfil$\m@th#1\updownarrow$\hfil
  }
}
\makeatother

\begin{document}
\title{Hierarchical Interpolative Factorization for Self Green's Function in 3D Modified Poisson-Boltzmann Equations}
\author{Yihui Tu$^1$, Zhenli Xu$^2$, Haizhao Yang$^3$  \\
$^1$ School of Mathematical Sciences, Shanghai Jiao Tong University, Shanghai 200240, China\\
$^2$ School of Mathematical Sciences, MOE-LSC, CMA-Shanghai and Shanghai Center\\ 
for Applied Mathematics, Shanghai Jiao Tong University, Shanghai, 200240, China\\
$^3$ Department of Mathematics, University of Maryland, College Park, MD, 20742, USA
}

\date{\today}
\maketitle

\abstract{
The modified Poisson-Boltzmann (MPB) equations are often used to describe equilibrium particle distribution of ionic systems. In this paper, we propose a fast algorithm to solve MPB equations with the self Green's function as the self energy in three dimensions, where the solution of the self Green's function poses a computational bottleneck due to the need to solve a high-dimensional partial differential equation.
Our algorithm combines the selected inversion with hierarchical interpolative factorization for the self Green's function by extending our previous result of two dimensions.
This leads to an $O(N\log N)$ algorithm through the strategical utilization of locality and low-rank characteristics of the corresponding operators. Furthermore, the estimated $O(N)$ complexity is obtained by applying cubic edge skeletonization at each level for thorough dimensionality reduction.
Extensive numerical results are performed to demonstrate the accuracy and efficiency of the proposed algorithm for problems in three dimensions.
}

\vspace{0.5cm}

\section{Introduction}

Electrostatic interaction plays important role in many systems at the nano-/microscale such as biomolecules, supercapacitors and charged soft matter \cite{Schoch:RMP:08,daiguji2004ion,BKN+:PR:2005,Liljestrm2015ElectrostaticSO}.
To provide a continuum description of charged systems, the Poisson-Boltzmann (PB) theory \cite{Gouy:JP:1910,Chapman:PM:1913} based on the mean-field assumption is a typical implicit solvent model describing the ionic distribution. It fails to account for many-body characteristics that are essential to describe electrostatic many-body behaviors of many systems, such as ion correlation and dielectric fluctuation. 

Various modified theories have been proposed \cite{modify1,Bazant2011DoubleLI, Eisenberg2020} to account for many-body effects, together with many numerical methods \cite{XML:PRE:2014,2014A,Liu2020APE}. 
The Gaussian variational field theory \cite{pnp1,Podgornik1989ElectrostaticCF} presents a promising approach to account for long-range Coulomb correlation, including dielectric variation \cite{pnp2,Liu2018ModifiedPM,Ma2020ModifiedPM}. This theory considers the self energy of a test ion as a correction to the mean-field potential energy, which is described by the self Green's function. 
By taking into account the self-energy correction, the effect of dielectric inhomogeneity can be incorporated \cite{modify3,Wang:PRE:2010,xu2,Ji2018AsymptoticAO}. 
The self Green's function used in the field theory satisfies the generalized Debye-H\"uckel (GDH) equation. The numerical solution of the GDH equation is computationally expensive due to its high spatial dimensions.
Based on the finite-difference discretization, the self Green's function corresponds to the diagonal of the inverse of the discrete elliptic differential operator of the GDH equation. The aim of our study is to calculate the self-energy in the GDH equation, a procedure that accelerates the numerical solution of the MPB equations, which requires efficient algorithm to determine the diagonal elements of the matrix inverse.

A straightforward method for extracting the diagonal of the matrix inverse is to first compute the entire matrix and then trivially extract the diagonal. This naive inversion approach has computational complexity of $O\left(N^{3}\right)$, which is the same as that of matrix factorization. In calculations of electronic structure and electrostatic correlation,  considerable effort is devoted to the development of an efficient method for acquiring the diagonal of the matrix inverse. A promising approach is the fast algorithm developed using sparsity and low-rankness. The selected inversion method was proposed by Lin {\it et al.}\cite{Lin2009,Lin2011:1,Lin2011:2} with $O(N^{3/2})$ computational complexity for 2D problems and  $O(N^{2})$ computational complexity for 3D problems, which involves a hierarchical decomposition of the computational domain $\boldsymbol{\Omega}$. The above method consists of two phases. Constructing the hierarchical Schur complements of the interior points for the blocks of the domain in a bottom-up pass, and then extracting the diagonal entries efficiently in a top-down pass by taking advantage of the hierarchical local dependence of the inverse matrices.
To further improve the efficiency of this method, Lin {\it et al.} \cite{Lin2011:1,Lin2011:2} exploited a supernode left-looking LDL factorization of the matrix, which significantly reduces the prefactor in computational complexity. Additionally, Xia {\it et al.} \cite{Xia2015} applied structured multifrontal LDL factorizations to achieve $O(N \text{poly}(\log N))$ complexity.

Recently, the hierarchical interpolative factorization (HIF) \cite{hifde,hifie} has been proposed to exploit a combination of multifrontal \cite{Brandt1977,Duff1983,George1973, Liu1992} and recursive dimensional reduction using frontal skeletonization. This approach aims to generate an approximate generalized LU/LDL decomposition with a linear or quasi-linear estimated computational cost. In contrast to previous methods 
\cite{Gillman2014ADS, Gillman2014AnOA, Grasedyck2007DomaindecompositionB, Schmitz2012AFD, Xia2009SuperfastMM}
that utilize fast structured methods to work implicitly with entire fronts while keeping them implicitly, the HIF offers the advantage of explicit front reduction. As a result, HIF significantly saves the resources needed to compute 3D problems and performs well on large-scale problems.

More recently, the selected inversion with the HIF (SelInvHIF) was proposed \cite{TU2022110893}.
The supernode left-looking LDL factorization is replaced with the HIF and the extraction phase is modified to approximate the diagonal of the inverse of the matrix within $O(N)$ operations for  2D problems. 
In this work, we further extend the SelInHIF to three-dimensional problems with $O(N \log N)$ complexity by face skeletonization and $O(N)$ complexity by means of skeletonizing cubic faces and then edges. 
For convenience, the former algorithm is still called the SelInvHIF and the latter is called ``SelInvHIF with edge skeletonization".
The computational complexity of the algorithm will be demonstrated through comprehensive theoretical derivation and the presentation of various numerical examples. In the subsequent section, the MPB is introduced, as it serves as an issue for testing the scaling of the algorithm within the context of three-dimensional problems.


The rest of the paper is organized as follows. 
Section \ref{SelInvHIF} introduces the skeletonization of matrix factorization and presents the SelInvHIF algorithm in detail.  
Then iterative solvers are described for MPB equations in Section \ref{NMmPBE}. 
In Section \ref{Numerical}, we present various numerical results obtained using the SelInvHIF algorithm.
Finally, we make the conclusion of the paper and discuss future work in Section \ref{Conclusion}.

\section{The SelInvHIF Algorithm}
\label{SelInvHIF}

Initially, we discuss details of SelInvHIF, followed by the introduction of SelInvHIF with edge skeletonization in Section \ref{SHIF-edge}. The SelInvHIF algorithm comprises two steps. In the first step, hierarchical Schur complements are constructed for the diagonal blocks of matrix $A$, which is discretized uniformly from the differential operator on a rectangular domain $\boldsymbol{\Omega}$. During the subsequent stage, the diagonal elements of $A^{-1}$ are extracted from the constructed hierarchy of Schur complements. 
Prior to the introduction of the formal description of the SelInvHIF algorithm,  we give a brief overview of the skeletonization of matrix factorization.

Let us determine some basic symbols and give necessary theorems for our algorithm. Given a matrix $A$, $A_{p q}$  or $A(I, J)$ is a submatrix with restricted rows and columns, where the \(p, q, r, I\) and \(J\) denote the ordered sets
of indices. For simplicity, matrix $A$ is assumed to be symmetric and nonsingular given by
\begin{equation}\numberwithin{equation}{section}
\label{eq:A}
A = \left[\begin{array}{ccc}
A_{pp} & A_{qp}^{T} &  \\
A_{qp} & A_{qq} & A_{rq}^{T} \\
 & A_{rq} & A_{rr}
\end{array}\right]
\end{equation}
which is defined over the indices \((p, q, r)\). In this matrix structure, \(p\) is related
to the degrees of freedom (DOFs) of the interior points on domain $\mathcal{D}$ ($\mathcal{D}$ is a subdomain of $\Omega$), $q$ to the DOFs on the
boundary \(\partial \mathcal{D}\), and \(r\) to the external domain \(\boldsymbol{\Omega} / \overline{\mathcal{D}}\). In general, the DOFs $q$ separates $p$ from $r$ which is often very large.  
Let  $G = A^{-1}$ and $G_{1} = A_{1}^{-1}$. $G_{pp}$ is the submatrix of $G$ corresponding to the row and column index set $p$, and  $A_{1}$ is the Schur complement of $A_{pp}$, i.e.,
\begin{equation*}
\label{eq:A1}
A_1=\left[\begin{array}{cc}
A_{qq}-A_{qp}A_{pp}^{-1}A_{qp}^{T} &  A_{rq}^{T}\\
A_{rq} & A_{rr}\\
\end{array}\right].
\end{equation*}

One of the preliminary tools used in the SelInvHIF is based on a crucial observation in the selected inversion method \cite{Lin2009}. Namely, to compute \(G_{p p}\), only the values of \(G_{1}\) with interaction in direct matrix $A$, \(\left(G_{1}\right)_{q q}\) are needed rather than the whole inverse of the Schur complement.
This implies that $G_{pp}$ is determined by $(G_{1})_{qq}$.
Furthermore, the diagonal entry $(G_{1})_{qq}$ can be calculated by utilizing a diagonal block of the inverse of the Schur complement of a submatrix of $A_1$. The recursive application of this approach leads to an efficient algorithm for computing $G_{pp}$. Specifically, we can compute a diagonal block of the $A^{-1}$ using a diagonal block of the inverse of a submatrix of $A$. Repeatedly applying this observation allows us to create an efficient recursive algorithm for computing the diagonal of $A^{-1}$. 

The interpolative decomposition (ID) \cite{Cheng:SIAM:05} for low-rank matrices based on Lemma \ref{lemma:2}  below is the second frequently used tool in the SelInvHIF. Suppose a disjoint partition of $q = \hat{q}\cup \check{q}$ with $|\hat{q}|=k$ is use. The sets $\hat{q}$ and $\check{q}$ are referred to as the skeleton and redundant indices, respectively. 

\begin{lemma}
\label{lemma:2}
Assume $A \in \mathbb{R}^{m\times n}$ with rank $k \leq \min(m,n)$ and $q$ be the set of all column indices of $A$. Then there exists a matrix $T_q \in \mathbb{R}^{k\times (n-k)}$ such that $A_{:,\check{q}} = A_{:,\hat{q}} T_{q}$.
\end{lemma}

Specifically, the redundant columns of matrix $A$ can be represented by the skeleton columns and the associated interpolation matrix from Lemma \ref{lemma:2}, and the following formula holds,

\begin{equation}
\label{cor}
A\left[\begin{array}{cc}
I & \\
-T_{q} & I\\
\end{array}\right]=
\left[\begin{array}{cc}A_{:,\check{q}} & A_{:,\hat{q}}\\
\end{array}\right]
\left[\begin{array}{cc}
I & \\
-T_{q} & I\\
\end{array}\right]
=
\left[\begin{array}{cc}
\boldsymbol{0} & A_{:,\hat{q}}\\
\end{array}\right].
\end{equation}
Eq.\eqref{cor} indicates that the sparsification of matrix $A$ is feasible by multiplying a triangular matrix which is formed from the interpolation matrix $T_q$ in Lemma \ref{lemma:2}.

The utilization of \eqref{cor} facilitates the elimination of redundant DOFs of a dense matrix featuring low-rank off-diagonal blocks, yielding a structured matrix of the form \eqref{eq:A}. This idea is referred to as block inversion with skeletonization and is reflected in Lemma \ref{lemma:4}. It is worth noting that the idea of skeletonization was first introduced in the HIF method \cite{hifde}.

\begin{lemma}
\label{lemma:4}
Let symmetric matrix $A$ have the following form

\begin{equation*}\numberwithin{equation}{section}
A =
\left[\begin{array}{cc}
A_{pp} & A_{qp}^{T}\\
A_{qp} & A_{qq}
\end{array}\right],
\end{equation*}
where $A_{qp}$ is numerically low-rank. Interpolative matrix $T_{p}$ satisfies $A_{q\check{p}} = A_{q\hat{p}} T_{p}$ with $p = \hat{p}\cup \check{p}$. Without loss of generality, rewrite

\begin{equation*}\numberwithin{equation}{section}
A = \left[\begin{array}{ccc}
A_{\check{p}\check{p}} & A_{\hat{p}\check{p}}^{T} & A_{q\check{p}}^{T} \\
A_{\hat{p}\check{p}} & A_{\hat{p}\hat{p}} & A_{q\hat{p}}^{T} \\
A_{q\check{p}} & A_{q\hat{p}} & A_{qq}
\end{array}\right]
\end{equation*}
and define

\begin{equation*}\numberwithin{equation}{section}
Q_{p} =
\left[\begin{array}{ccc}
I & & \\
-T_{p} & I & \\
& & I
\end{array}\right].
\end{equation*}
Let $\bar{A} \triangleq Q_{p}^{T}A Q_{p}$. Then one has

\begin{equation}\label{eq:Ab}
\numberwithin{equation}{section}
\bar{A}  =
\left[\begin{array}{ccc}
B_{\check{p}\check{p}} & B_{\hat{p}\check{p}}^{T} &  \\
B_{\hat{p}\check{p}} & A_{\hat{p}\hat{p}} & A_{q\hat{p}}^{T} \\
 & A_{q\hat{p}} & A_{qq}
\end{array}\right],
\end{equation}
with $B_{\check{p}\check{p}} = A_{\check{p}\check{p}}-T_{p}^{T}A_{\hat{p}\check{p}}-A_{\hat{p}\check{p}}^{T}T_{p}+T_{p}^{T}A_{\hat{p}\hat{p}}T_{p}$, and  $B_{\hat{p}\check{p}} = A_{\hat{p}\check{p}} - A_{\hat{p}\hat{p}}T_{p}.$

Further suppose that $B_{\check{p}\check{p}}$ is nonsingular. Let $G = A^{-1}$, $\bar{G} = \bar{A}^{-1}$, $G_{1} = G_{\hat{p}\cup q, \hat{p}\cup q}$, and $\bar{A}_{1}$ be the Schur complement of  $B_{\check{p}\check{p}}$, i.e.,

\begin{equation*}\numberwithin{equation}{section}
\label{eq:Schur1}
\bar{A}_{1}=
\left[\begin{array}{cc}
A_{\hat{p}\hat{p}}-B_{\hat{p}\check{p}}B_{\check{p}\check{p}}^{-1}B_{\hat{p}\check{p}}^{T} &  A_{q\hat{p}}^{T}\\
A_{q\hat{p}} & A_{qq}\\
\end{array}\right],
\end{equation*}
and $\bar{G}_{1} = \bar{A}_{1}^{-1}$. Then, by Eq. \eqref{eq:Ab} the following formulas holds,

\begin{equation*}\numberwithin{equation}{section}
G_{\check{p}\check{p}} = \bar{G}_{\check{p}\check{p}} = B_{\check{p}\check{p}}^{-1}+\begin{bmatrix} -B_{\check{p}\check{p}}^{-1}B_{\hat{p}\check{p}}^{T}\ \boldsymbol{0} \end{bmatrix}\bar{G}_{1}\begin{bmatrix} -B_{\check{p}\check{p}}^{-1}B_{\hat{p}\check{p}}^{T}\ \boldsymbol{0} \end{bmatrix}^T,
\end{equation*}

\begin{equation*}\numberwithin{equation}{section}
G_{1} =
\left[\begin{array}{cc}
T_{p}B_{\check{p}\check{p}}^{-1}T_{p}^{T} & \\
& \boldsymbol{0}
\end{array}\right]+
\left[\begin{array}{cc}
T_{p}B_{\check{p}\check{p}}^{-1}B_{\hat{p}\check{p}}^{T}+I & \\
& I
\end{array}\right]\bar{G}_{1}
\left[\begin{array}{cc}
B_{\hat{p}\check{p}}B_{\check{p}\check{p}}^{-1}T_{p}^{T}+I & \\
& I
\end{array}\right].
\end{equation*}
 \end{lemma}
 
Lemma  \ref{lemma:4} shows that computing $G_{\check{p}\check{p}}$ requires only the values of $\bar{G}_1$ associated with row and column indices in $\hat{p}$, rather than the whole inverse of the Schur complement. Thus, $G_{\check{p}\check{p}}$ is determined by $(\bar{G}_1)_{\hat{p}\hat{p}}$, a diagonal block of $\bar{A}_1^{-1}$, which has a smaller size than the original matrix $A$. Although $\bar{A}_1$ may be dense, if it has low-rank off-diagonal blocks, then the same approach used in Eq. \eqref{eq:Ab} can be applied to compute a diagonal block of $\bar{G}_1$, resulting in a recursive algorithm that efficiently computes the diagonal blocks of $G$.

This skeletonization technique was proposed by Ho and Ying \cite{hifde} and is based on the observation that the Schur complements have specific low-rank structures.  Specifically, $A_{pp}^{-1}$, obtained from a local differential operator, often has low-rank off-diagonal blocks. Additionally, numerical experiments illustrate that the Schur complement interaction $A_{qq}-A_{qp}A_{pp}^{-1}A_{qp}^{T}$ also possesses the same rank structure. In the following subsection, we use Lemma \ref{lemma:4} to generate hierarchical Schur complements for diagonal blocks of $A$.

\subsection{Hierarchy of Schur complements}
\label{21}

To achieve a hierarchical disjoint partition for the differential operator in domain $\boldsymbol{\Omega}$, bipartitioning is performed in each dimension, resulting in leaf domains of size $r_0\times r_0\times r_0$ and a total integer level $L$.
Domain $\boldsymbol{\Omega}$ is defined by a grid size of $\sqrt[3]{N}\times \sqrt[3]{N}\times \sqrt[3]{N}= r_0 2^{L-1}\times r_0 2^{L-1}\times r_0 2^{L-1}$ and is associated with a matrix $A$ of size $N\times N$. 
Furthermore, to take advantage of the low-rankness of matrix $A$, $L-1$ fractional levels are introduced between $L$ integer levels.
The hierarchy construction of Schur complements is carried out at levels $1$, $3/2$, $2$, $5/2$, $\dots$, and $L$.

\begin{figure}[H]
	\centering
	\includegraphics[width=0.8\textwidth]{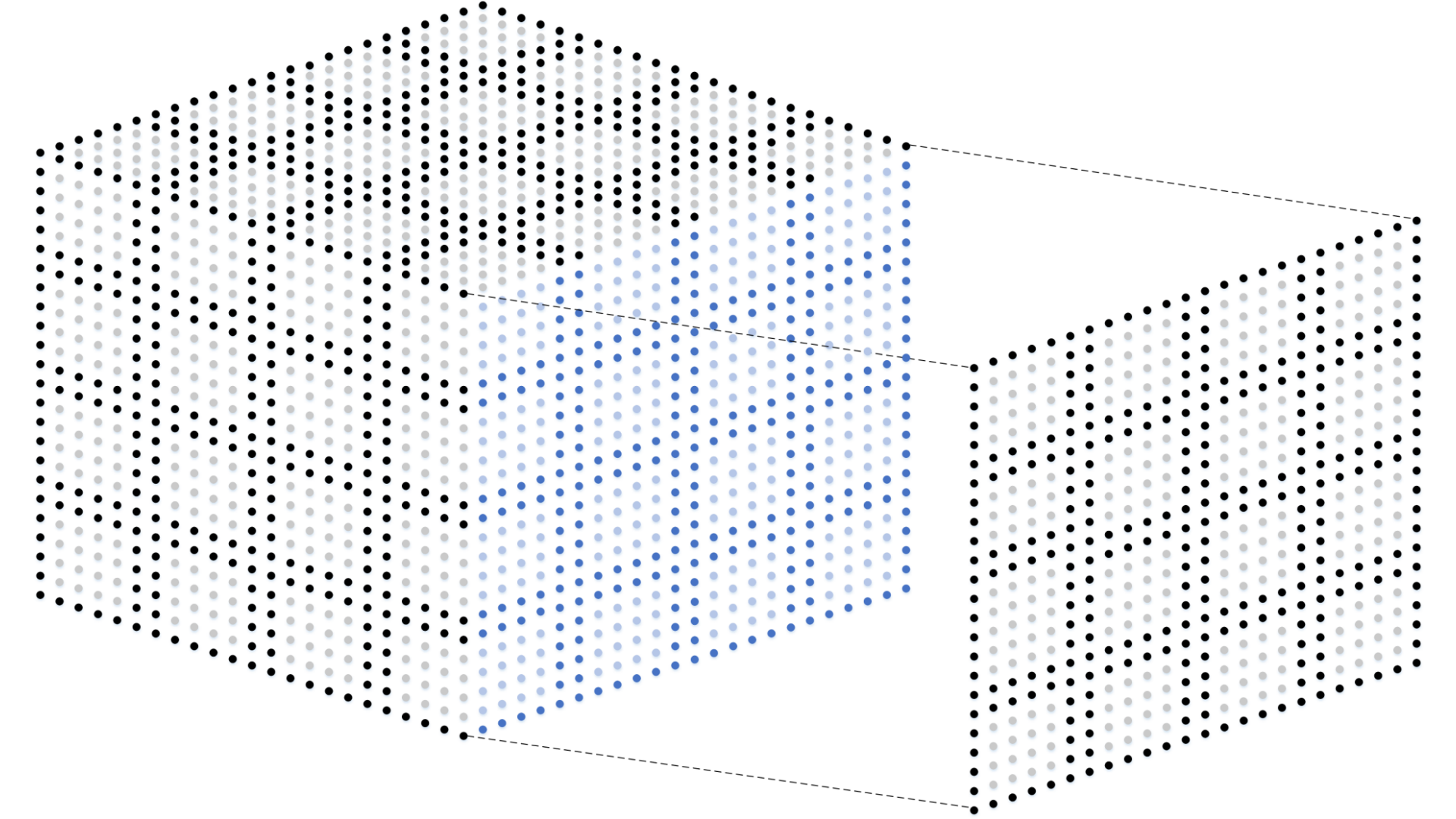}
	\caption{The DOFs in the bottom level. In this level, the domain is divided into 64 blocks.
The interior points are indicated by gray (or light blue), while the boundary points are indicated by black (or blue) in this and following integer levels. Note that the prefactor is reduced due to share faces (edges).}
        \label{fig:lel1-0}
\end{figure}

Let us consider the case of $r_0=6$ and $L=3$ to describe the process in detail without loss of generality.
Initially, the entire domain is regarded as the top level (Level $3$) and is partitioned into eight blocks at the next level (Level $2$).  Each block is further partitioned into eight sub-blocks at a lower level (Level $1$), resulting in a total of $2^{L-1}\times 2^{L-1}\times 2^{L-1}=64$ blocks at the bottom level, as illustrated in Figure \ref{fig:lel1-0}. In addition,
one fractional level is considered between two consecutive integer levels and the low-rank matrices that represent the fronts between domain blocks are reduced into skeletons by this level.

\subsubsection{Bottom level  \texorpdfstring{$\bm{\ell=1}$}{}} 

The initial index set $J_{0}$ follows the row-major ordering, while domain $\boldsymbol{\Omega}$ is hierarchically partitioned into disjoint blocks at level $\ell=1$, with each block having a size of $2^{L-\ell}\times 2^{L-\ell}\times 2^{L-\ell}=4\times 4\times 4$. 
All points within each block are classified into interior and boundary points, where the former are not related to the points in other blocks, and the latter are related to the neighboring points in other blocks. 
The interior points are denoted as $I_{1;ijk}$ (shown in gray or light blue in Figure \ref{fig:lel1-0}) and the boundary points are denoted as $J_{1;ijk}$ (shown in black or blue in Figure \ref{fig:lel1-0}) for each block, where $i,j,k=1,2,3,4$ are the indices of the blocks in each dimension. 
The differential operators have a locality property, which implies that $A(I_{1;ijk},I_{1;i'j'k'}) = 0$ (or $A(I_{1;ijk},J_{1;i'j'k'}) = 0$) if $(i,j,k) \neq (i',j',k')$.

The interior points are removed using block inversion. One can then focuses the problem on the boundary points. 
To achieve this, one uses the proper row and column permutations to matrix $A$ defined with the index set $J_ 0$ in order to place all of the interior points in front of the boundary points. 
Actually, matrix $A$ can be permuted into a new matrix by a permutation matrix $P_1$ as follows,

\begin{equation}\numberwithin{equation}{section}
\label{A1}
A_{1} = P_{1}^{-1}A P_{1}=
\left[\begin{array}{cc}
U_{1} & V_{1}^{T}\\
V_{1} & W_{1}
\end{array}\right]
\end{equation}
with index set $(I_{1}|J_{1})$, $U_{1} = A_{1}(I_{1},I_{1})$, $V_{1} = A_{1}(J_{1},I_{1})$, and $W_{1} = A_{1}(J_{1},J_{1})$. Here $I_{1}$ represents the indices of all interior points, denoted as $I_{1} = I_{1;111}I_{1;121}...I_{1;444}$, and $J_{1}$ represents the indices of all boundary points, denoted as $J_{1}=J_{1;111}J_{1;121}...J_{1;444}$. 

Due to the locality property, both $U_1$ and $V_1$ are block diagonal matrices. Figure \ref{fig:lel1-0} shows that interior points in different blocks are not connected. Boundary points in each block are only connected to the interior points in the same block. Furthermore, $U_1$ and $V_1$ are of the following form,

\begin{equation*}\numberwithin{equation}{section}
U_{1} =
\left[\begin{array}{cccc}
U_{1;111}& & & \\
& U_{1;121} & & \\
& & \ddots & \\
& & & U_{1;444}
\end{array}\right],
V_{1} =
\left[\begin{array}{cccc}
V_{1;111} & & & \\
& V_{1;121} & & \\
& & \ddots &  \\
& & & V_{1;444}
\end{array}\right],
\end{equation*}

with $U_{1;ijk} = A_{1}(I_{1;ijk},I_{1;ijk})$ and $V_{1;ijk} = A_{1}(J_{1;ijk},I_{1;ijk})$, for $i,j,k=1,2,3,4$.

Using Gaussian elimination, one can obtain,
\begin{equation}\numberwithin{equation}{section}
\label{eq:A1-1}
A_{1}^{-1} =
L_{1}^{T}
\left[\begin{array}{cc}
U_{1}^{-1} & \\
 & (W_{1}-V_{1}U_{1}^{-1}V_{1}^{T})^{-1}
\end{array}\right]L_{1},
\text{with} \quad
L_{1} =
\left[\begin{array}{cc}
I & \\
-V_{1}U_{1}^{-1} & I
\end{array}\right].
\end{equation}
Since $U_1$ is a block diagonal matrix with each diagonal block of a size $(r_0-2)^3 \times (r_0-2)^3$, its inverse can be computed directly.
By using the block diagonal matrices $V_{1}$ and $U_{1}^{-1}$, $V_{1}U_{1}^{-1} $ is also a block diagonal matrix and can be computed independently within each block,
\begin{equation*}\numberwithin{equation}{section}
V_{1}U_{1}^{-1} =
\left[\begin{array}{cccc}
V_{1;111}U_{1;111}^{-1} & & & \\
& V_{1;121}U_{1;121}^{-1} & & \\
& & \ddots & \\
& & & V_{1;444}U_{1;444}^{-1}
\end{array}\right].
\end{equation*}
Similarly, the block diagonal matrix $V_{1}U_{1}^{-1}V_{1}^{T}$ is expressed as

\begin{equation*}\numberwithin{equation}{section}
V_{1}U_{1}^{-1}V_{1}^{T} =
\left[\begin{array}{cccc}
V_{1;111}U_{1;111}^{-1}V_{1;111}^{T} & & & \\
& V_{1;121}U_{1;121}^{-1}V_{1;121}^{T} & & \\
& & \ddots & \\
& & & V_{1;444}U_{1;444}^{-1}V_{1;444}^{T}
\end{array}\right].
\end{equation*}
Combining (\ref{A1}) and (\ref{eq:A1-1}), one has 

\begin{equation}\numberwithin{equation}{section}
G  =  P_{1}A_{1}^{-1}P_{1}^{-1} = P_{1}L_{1}^{T}
\left[\begin{array}{cc}
U_{1}^{-1} & \\
& G_{1}
\end{array}\right]L_{1}P_{1}^{-1},
\label{eqn:G1}
\end{equation}
where $G_1={(W_{1}-V_{1}U_{1}^{-1}V_{1}^{T})}^{-1}$ is the inverse of the Schur complement of $U_1$. Consequently, by removing interior points from matrix $A$, one is able to simplify the problem.
\subsubsection{Fractional level  \texorpdfstring{$\bm{\ell=3/2}$}{}} 

\label{sec:schur3/2}

\begin{figure}[H]
	\centering
	\includegraphics[width=0.4\textwidth]{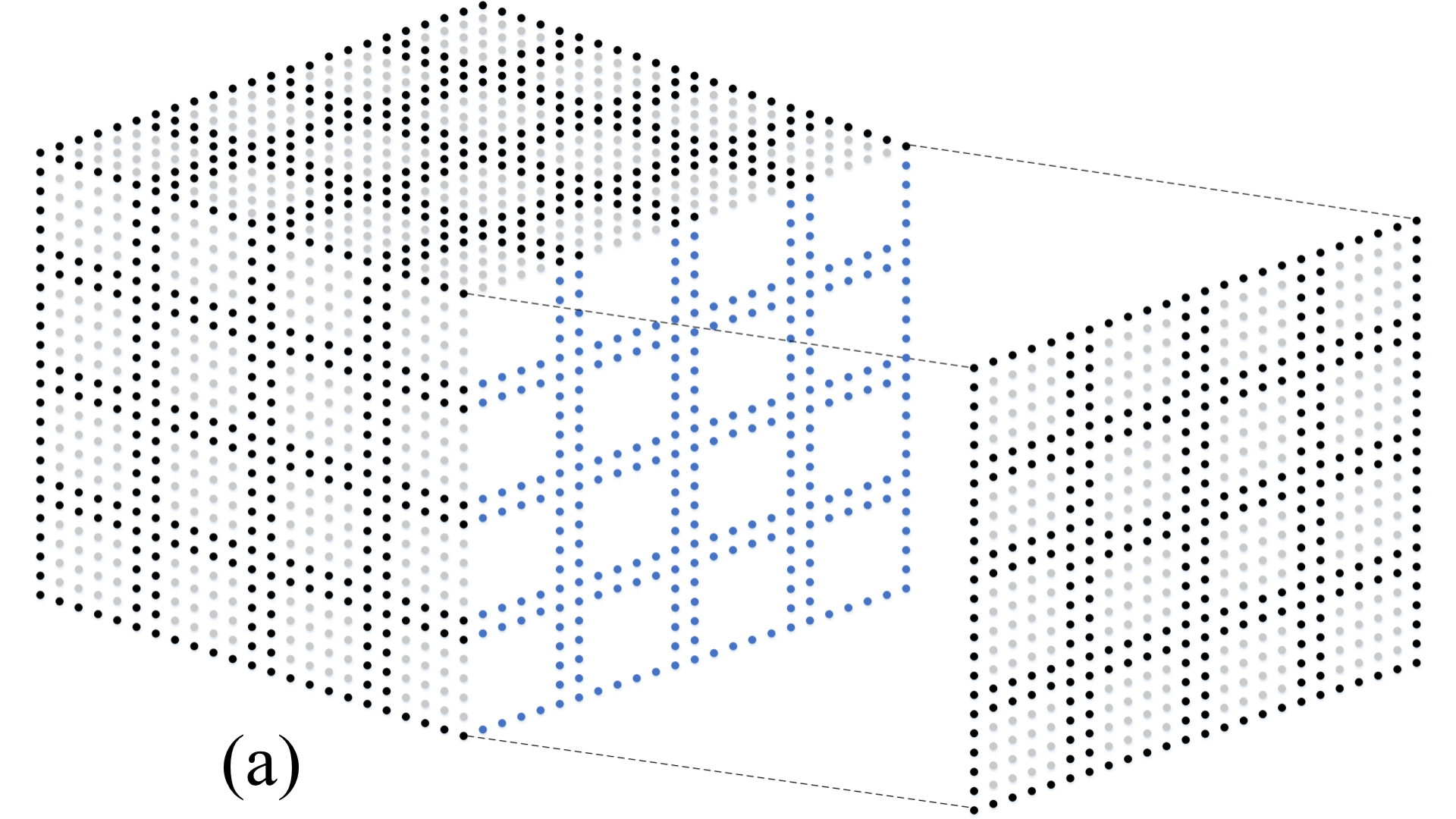}
	\includegraphics[width=0.4\textwidth]{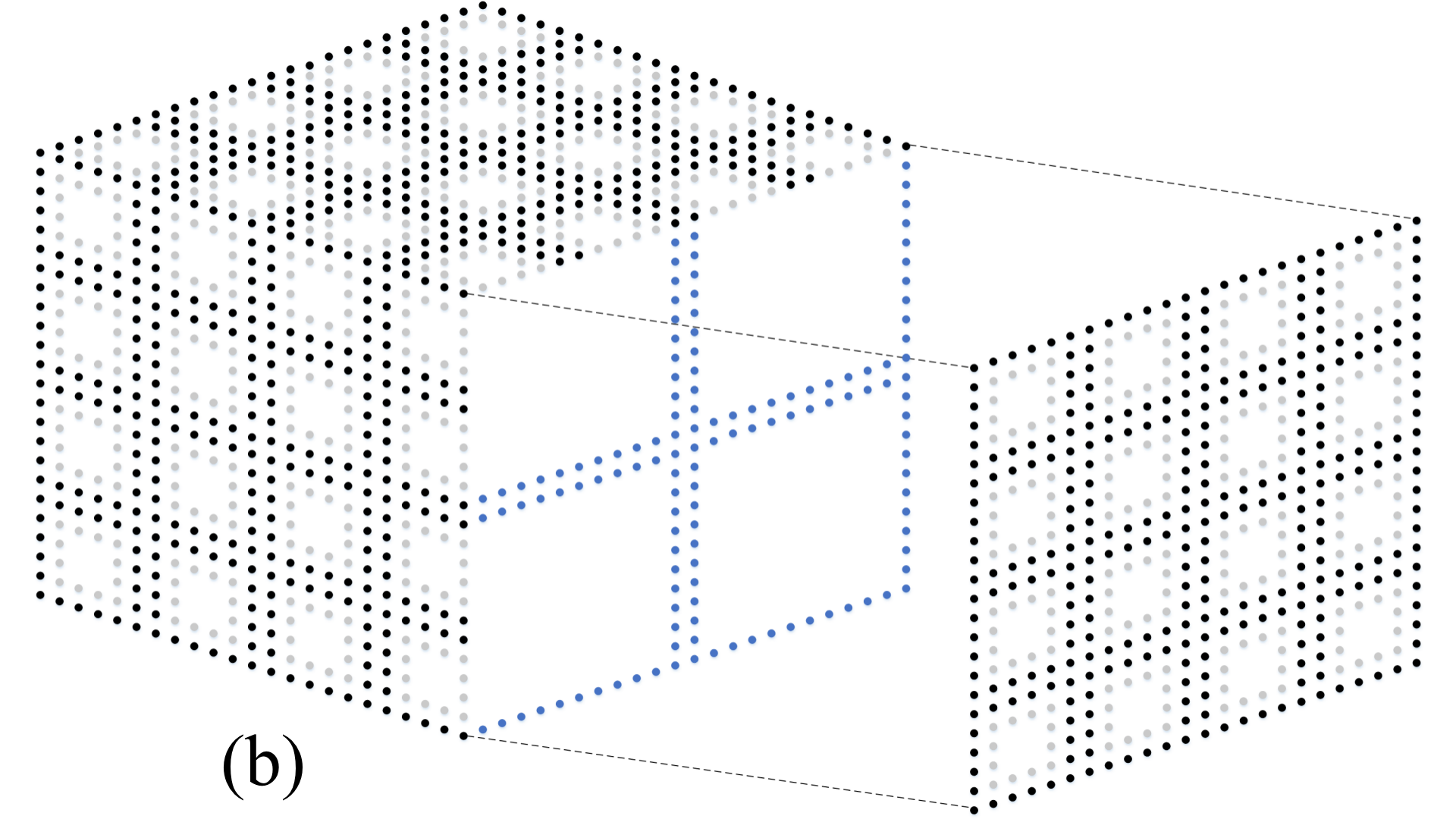}
	\includegraphics[width=0.4\textwidth]{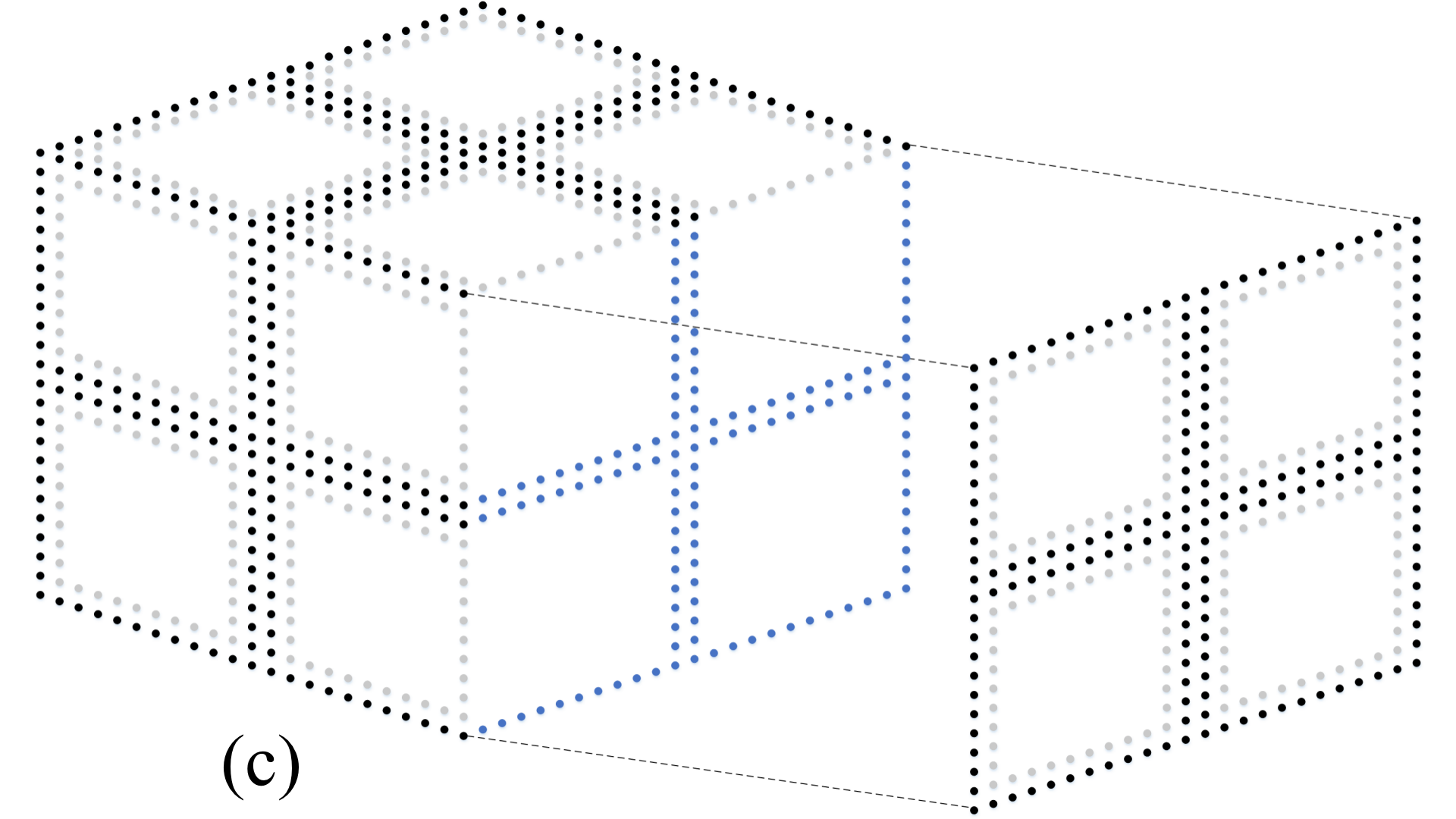}
	\includegraphics[width=0.4\textwidth]{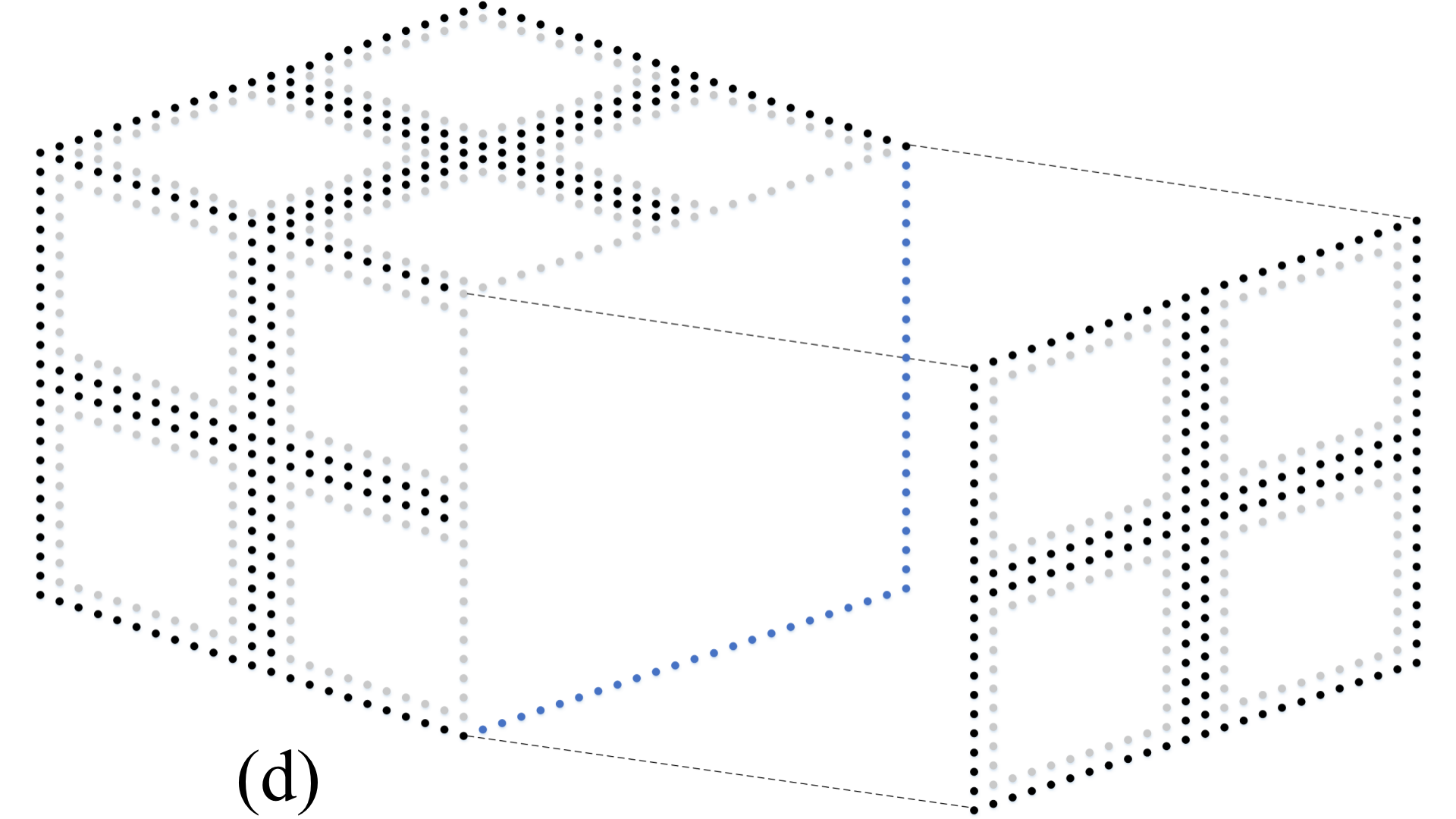}
	\caption{(a): The DOFs in the level $\ell=3/2$. The domain is divided into 384 faces of 64 blocks.
The skeleton points are indicated in black (or light blue), and the redundant points are indicated in gray. (b): The DOFs in the level $\ell=2$ after elimination. The domain is divided into 8 blocks. (c): The DOFs in the level $\ell=5/2$ after skeletonization. (d): The DOFs in the level $\ell=3$. This is the top level.}
        \label{fig:lel2-0}
\end{figure}

At this level, the aim is to obtain $G_{1}$ in \eqref{eqn:G1} on index set $J_{1}$, which corresponds to the boundary points of the domain blocks at the first level. 
Domain $\boldsymbol{\Omega}$ is partitioned into 64 blocks and the total number of faces in blocks is 384 in this example with $L=3$ (see Figure \ref{fig:lel2-0} (a)).
Each face consists of  the DOFs inside the corresponding area and some of the DOFs on its boundary. In addition, a face not only interacts within its own block but also interacts with faces in neighbor blocks. The associated matrix permits low-rank off-diagonal blocks since the DOFs of a face only interact with a few number of other neighboring blocks.  Furthermore, Lemma \ref{lemma:4} is applied to skeletonize the DOFs on the faces in each block. An ID can be implemented to select the redundant and skeleton DOFs approximately in each block, and the interpolation matrix $T_{q}$ can be recorded as in Lemma \ref{lemma:2}. 
Figure \ref{fig:lel2-0} (a) shows the $i$th face, where the redundant DOFs are identified by $I_{3/2;i}$, the skeleton DOFs , by $J_{3/2;i}$, and the associated interpolation matrix, by $T_{3/2;i}$.

Similar to the bottom level, an appropriate permutation matrix $P_{3/2}$ is designed in order to move all of the redundant points in front of the skeleton points and reindex $J_{1}$ by the permutation matrix. Furthermore, matrix 
$W_{1}-V_{1}U_{1}^{-1}V_{1}^{T}$ can be permuted into a new matrix by the permutation matrix $P_{3/2}$  as follows,

\begin{equation*}\numberwithin{equation}{section}
A_{\frac{3}{2}} = P_{\frac{3}{2}}^{-1}(W_{1}-V_{1}U_{1}^{-1}V_{1}^{T})P_{\frac{3}{2}} =
\left[\begin{array}{cc}
U_{\frac{3}{2}} & V_{\frac{3}{2}}^{T} \\
V_{\frac{3}{2}} & W_{\frac{3}{2}}
\end{array}\right]
\end{equation*}
with index set $(I_{3/2}|J_{3/2})$,
$U_{3/2} = A_{3/2}(I_{3/2},I_{3/2}), V_{3/2} = A_{3/2}(J_{3/2},I_{3/2})$,  and the dense matrix\\
 $W_{3/2} = A_{3/2}(J_{3/2},J_{3/2})$. Here $I_{3/2}$ represents the indices of all redundant  points, denoted as $I_{3/2;1}I_{3/2;2}...I_{3/2;384}$, and $J_{3/2}$ represents the indices of all skeleton points, such that it  can be denoted as $J_{3/2;1}J_{3/2;2}...J_{3/2;384}$. 

Denote $T_{3/2}$ by a block diagonal matrix

\begin{equation*}\numberwithin{equation}{section}
\label{eq:T3/2}
T_{\frac{3}{2}} =
\left[\begin{array}{ccc}
-T_{\frac{3}{2};1} & & \\
& \ddots & \\
& & -T_{\frac{3}{2};384}
\end{array}\right],
\end{equation*}
and arrange a $|J_{1}|\times |J_{1}|$ matrix

\begin{equation*}\numberwithin{equation}{section}
\label{eq:Q3/2}
Q_{\frac{3}{2}} =
\left[\begin{array}{cc}
I &  \\
T_{\frac{3}{2}}& I  \\
\end{array}\right].
\end{equation*}
Thus, the new matrix is updated

\begin{equation*}\numberwithin{equation}{section}
\bar{A}_{\frac{3}{2}} = Q_{\frac{3}{2}}^{T}A_{\frac{3}{2}}Q_{\frac{3}{2}} =
\left[\begin{array}{cc}
\bar{U}_{\frac{3}{2}} & \bar{V}_{\frac{3}{2}}^{T} \\
\bar{V}_{\frac{3}{2}} & W_{\frac{3}{2}}
\end{array}\right],
\end{equation*}
where $\bar{U}_{3/2}$ and $\bar{V}_{3/2}$ are block diagonal matrices with

\begin{equation*}\numberwithin{equation}{section}
\bar{U}_{\frac{3}{2}}(I_{\frac{3}{2};i},I_{\frac{3}{2};j}) = 0,\quad  \bar{V}_{\frac{3}{2}}(J_{\frac{3}{2};i},I_{\frac{3}{2};j}) = 0, \quad\forall i \neq j.
\end{equation*}
Similarly, one can obtain the following inverse by Gaussian elimination,

\begin{equation*}\numberwithin{equation}{section}
\bar{A}_{\frac{3}{2}}^{-1}  =
L_{\frac{3}{2}}^{T}
\left[\begin{array}{cc}
\bar{U}_{\frac{3}{2}}^{-1} &  \\
 & G_{\frac{3}{2}}
\end{array}\right]L_{\frac{3}{2}}
\end{equation*}
with

\begin{equation*}\numberwithin{equation}{section}
L_{\frac{3}{2}}=
\left[\begin{array}{cc}
I &  \\
-\bar{V}_{\frac{3}{2}}\bar{U}_{\frac{3}{2}}^{-1} & I
\end{array}\right],\quad
G_{\frac{3}{2}} = (W_{\frac{3}{2}}-\bar{V}_{\frac{3}{2}}\bar{U}_{\frac{3}{2}}^{-1}\bar{V}_{\frac{3}{2}}^{T})^{-1}
\end{equation*}
as in Lemma \ref{lemma:4}.  Since, $-\bar{V}_{3/2}\bar{U}_{3/2}^{-1}$ and $\bar{V}_{3/2}\bar{U}_{3/2}^{-1}\bar{V}_{3/2}^{T}$ are block diagonal matrices, they can be computed independently within each block. Thus, 

\begin{equation*}\numberwithin{equation}{section}
G_{1} \approx P_{\frac{3}{2}}Q_{\frac{3}{2}}L_{\frac{3}{2}}^{T}
\left[\begin{array}{cc}
\bar{U}_{\frac{3}{2}}^{-1} &  \\
 & G_{\frac{3}{2}}
\end{array}\right]L_{\frac{3}{2}}Q_{\frac{3}{2}}^{T}P_{\frac{3}{2}}^{-1}.
\end{equation*}
Therefore, the inversion problem is reduced to a smaller matrix  $W_{3/2}-\bar{V}_{3/2}\bar{U}_{3/2}^{-1}\bar{V}_{3/2}^{T}$ by eliminating the redundant DOFs as in Lemma \ref{lemma:4}.

\subsubsection{Middle level  \texorpdfstring{$\bm{\ell=2}$}{}}

At Level $\ell=2$, the domain $\boldsymbol{\Omega}$ consists of $2^{L-\ell}\times 2^{L-\ell}\times 2^{L-\ell}=2\times 2\times2$ blocks with interior and boundary points. Similar to the former integer level, permutation matrix $P_{2}$ is applied to reindex the points in $J_{3/2}$ into $I_{2}$ and $J_{2}$,
\begin{equation*}\numberwithin{equation}{section}
J_{\frac{3}{2}} \stackrel{P_{2}}{\longrightarrow} (I_{2;11}I_{2;12}I_{2;21}I_{2;22}|J_{2;11}J_{2;12}J_{2;21}J_{2;22}):=(I_{2}|J_{2}).
\end{equation*}
Use the same strategy as at Level $1$ and denote

\begin{equation*}\numberwithin{equation}{section}
A_{2} = P_{2}^{-1}(W_{\frac{3}{2}}-\bar{V}_{\frac{3}{2}}\bar{U}_{\frac{3}{2}}^{-1}\bar{V}_{\frac{3}{2}}^{T})P_{2} =
\left[\begin{array}{cc}
U_{2} & V_{2}^{T} \\
V_{2} & W_{2}
\end{array}\right]
\end{equation*}
with

\begin{equation*}\numberwithin{equation}{section}
U_{2} = A_{2}(I_{2},I_{2}),\quad V_{2} = A_{2}(J_{2},I_{2}), \quad\text{and}\quad W_{2} = A_{2}(J_{2},J_{2}).
\end{equation*}
It can be observed that matrices $U_2$ and $V_2$ possess a block diagonal structure. Thus,

\begin{equation*}\numberwithin{equation}{section}
G_{\frac{3}{2}} = P_{2}L_{2}^{T}
\left[\begin{array}{cc}
U_{2}^{-1} &  \\
 & G_{2}
\end{array}\right]L_{2}P_{2}^{-1},
\end{equation*}
and

\begin{equation*}\numberwithin{equation}{section}
L_{2} =
\left[\begin{array}{cc}
I &  \\
-V_{2}U_{2}^{-1} & I
\end{array}\right], \text{and} \quad 
G_{2} = (W_{2}-V_{2}U_{2}^{-1}V_{2}^{T})^{-1}.
\end{equation*}
Finally,  one is able to simplify the problem to smaller matrix $W_{2}-V_{2}U_{2}^{-1}V_{2}^{T}$ by removing interior points. The DOFs are shown in Figure \ref{fig:lel2-0} (b) after elimination in level $\ell=2$. 
\subsubsection{Fractional  \texorpdfstring{$\bm{\ell=5/2}$}{}} 

As in $\ell=3/2$, at this level, one aims  to find $G_{2}$ indexed by $J_{2}$. Again, one divides domain $\boldsymbol{\Omega}$ into 8 blocks with 48 faces. Through the ID, one distinguishes the redundant DOFs $I_{5/2;i}$ and the skeleton DOFs $J_{5/2;i}$ in the $i$th face, and records the interpolation matrix $T_{5/2;i}$. Reindex $J_{2}$ with a permutation matrix $P_{5/2}$ such that

\begin{equation*}\numberwithin{equation}{section}
J_{2} \stackrel{P_{\frac{5}{2}}}{\longrightarrow} (I_{\frac{5}{2};1}I_{\frac{5}{2};2}I_{5/2;3}I_{\frac{5}{2};48}|J_{\frac{5}{2};1}J_{\frac{5}{2};2}J_{\frac{5}{2};3}J_{\frac{5}{2};48}):=(I_{\frac{5}{2}}|J_{\frac{5}{2}}).
\end{equation*}
Denote

\begin{equation*}\numberwithin{equation}{section}
\label{eq:T5/2}
T_{\frac{5}{2}} =
\left[\begin{array}{ccc}
-T_{\frac{5}{2};1} & & \\
& \ddots & \\
& & -T_{\frac{5}{2};48}
\end{array}\right]
\end{equation*}
and a $|J_{2}|\times |J_{2}|$ matrix

\begin{equation*}\numberwithin{equation}{section}
\label{eq:Q5/2}
Q_{\frac{5}{2}} =
\left[\begin{array}{cc}
I &  \\
T_{\frac{5}{2}}& I \\
\end{array}\right].
\end{equation*}
Then

\begin{equation*}\numberwithin{equation}{section}
\bar{A}_{\frac{5}{2}} = Q_{\frac{5}{2}}^{T}P_{\frac{5}{2}}^{-1}(W_{2}-V_{2}U_{2}^{-1}V_{2}^{T})P_{\frac{5}{2}}Q_{\frac{5}{2}} =
\left[\begin{array}{cc}
\bar{U}_{\frac{5}{2}} & \bar{V}_{\frac{5}{2}}^{T} \\
\bar{V}_{\frac{5}{2}} & W_{\frac{5}{2}}
\end{array}\right]
\end{equation*}
with

\begin{equation*}\numberwithin{equation}{section}
\bar{U}_{\frac{5}{2}}(I_{\frac{5}{2};i},I_{\frac{5}{2};j}) = 0, \quad \bar{V}_{\frac{5}{2}}(J_{\frac{5}{2};i},I_{\frac{5}{2};j}) = 0,\quad  \forall i \neq j.
\end{equation*}
Therefore,

\begin{equation*}\numberwithin{equation}{section}
G_{2} \approx P_{\frac{5}{2}}Q_{\frac{5}{2}}L_{\frac{5}{2}}^{T}
\left[\begin{array}{cc}
\bar{U}_{\frac{5}{2}}^{-1} &  \\
 & G_{\frac{5}{2}}
\end{array}\right]L_{\frac{5}{2}}Q_{\frac{5}{2}}^{T}P_{\frac{5}{2}}^{-1},
\end{equation*}
with

\begin{equation*}\numberwithin{equation}{section}
L_{\frac{5}{2}} =
\left[\begin{array}{cc}
I &  \\
-\bar{V}_{\frac{5}{2}}\bar{U}_{\frac{5}{2}}^{-1} & I
\end{array}\right], \text{and}\quad
G_{\frac{5}{2}} = (W_{\frac{5}{2}}-\bar{V}_{\frac{5}{2}}\bar{U}_{\frac{5}{2}}^{-1}\bar{V}_{\frac{5}{2}}^{T})^{-1}.
\end{equation*}
Note that $\bar{U}_{5/2}$ and $\bar{V}_{5/2}$ are block diagonal.
The matrix inversion problem now has been reduced to $W_{5/2}-\bar{V}_{5/2}\bar{U}_{5/2}^{-1}\bar{V}_{5/2}^{T}$. The DOFs are shown in Figure \ref{fig:lel2-0} (c) after skeletonization in level $\ell=5/2$. 

\subsubsection{Top level  \texorpdfstring{$\bm{\ell=3}$}{}}

Partition domain $\boldsymbol{\Omega}$ into $2^{L-\ell}\times 2^{L-\ell}\times 2^{L-\ell}=1\times 1\times 1 $ block (i.e., no partition at this level, in Figure \ref{fig:lel2-0} (d).). Similarly, the index set $J_{5/2}$ is reindexed by partitioning it into the union of an interiors index set $I_3$ and a boundary index set $J_3$, using a permutation matrix $P_3$ as follows:

\begin{equation*}\numberwithin{equation}{section}
J_{\frac{5}{2}} \stackrel{P_{3}}{\longrightarrow} (I_{3}|J_{3}).
\end{equation*}
Thus, one has
\begin{equation*}\numberwithin{equation}{section}
G_{\frac{5}{2}} = P_{3}L_{3}^{T}
\left[\begin{array}{cc}
U_{3}^{-1} &  \\
 & G_{3}
\end{array}\right]L_{3}P_{3}^{-1},
\end{equation*}
with

\begin{equation*}\numberwithin{equation}{section}
L_{3} =
\left[\begin{array}{cc}
I &  \\
-V_{3}U_{3}^{-1} & I
\end{array}\right],\text{and} \quad
G_{3} = (W_{3}-V_{3}U_{3}^{-1}V_{3}^{T})^{-1}.
\end{equation*}
In this top level, the inverse of $G_{3}$ can be computed directly because of its small size.


\subsubsection{The algorithm for the hierarchy of Schur complements}

In this section, we aim to construct a hierarchical structure of Schur complements for matrix $A$ is constructed on an $\sqrt[3]{N} \times \sqrt[3]{N} \times \sqrt[3]{N}$ grid. The process involves dividing the points in each block at each integer level into interior and boundary points. Specifically, the interior points are only involved in interactions with other points within the same block, prompting a reindexing and subsequent elimination of the interior points. At each fractional level, face skeletonization is considered, and an ID approach is applied to distinguish redundant and skeleton points. Here, the redundant points only interact with other points in the same cell, leading to a reindexing and elimination of the redundant points.

The relationships between levels are defined as follows,

\begin{equation}\numberwithin{equation}{section}
\label{eq:rel}
G_{\ell} =
\begin{cases}
 A^{-1}, & \ell = 0; \\
(W_{\ell}-V_{\ell}U_{\ell}^{-1}V_{\ell}^{T})^{-1}, & \ell \text{\ is\ integer}; \\
(W_{\ell}-\bar{V}_{\ell}\bar{U}_{\ell}^{-1}\bar{V}_{\ell}^{T})^{-1}, & \ell \text{\ is\ fractional}.
\end{cases}
\end{equation}
Based on (\ref{eq:rel}), it follows the recursive relation with integer $\ell$,

\begin{equation*}\numberwithin{equation}{section}
G_{\ell-1} \approx P_{\ell-\frac{1}{2}}Q_{\ell-\frac{1}{2}}L_{\ell-\frac{1}{2}}^{T}
\left[\begin{array}{cc}
\bar{U}_{\ell-\frac{1}{2}}^{-1} &  \\
 & G_{\ell-\frac{1}{2}}
\end{array}\right]L_{\ell-\frac{1}{2}}Q_{\ell-\frac{1}{2}}^{T}P_{\ell-\frac{1}{2}}^{-1},
\end{equation*}
\begin{equation*}\numberwithin{equation}{section}
G_{\ell-\frac{1}{2}} = P_{\ell}L_{\ell}^{T}
\left[\begin{array}{cc}
U_{\ell}^{-1} &  \\
 & G_{\ell}
\end{array}\right]L_{\ell}P_{\ell}^{-1}.
\end{equation*}
Therefore, the hierarchy of Schur complements  can be constructed from Level 1. One describes the steps in Algorithm \ref{algo1}. Note that the reindexing is implicitly included in Algorithm \ref{algo1}, when one uses the index sets $I_{\ell;ijk}$ and $J_{\ell;ijk}$ or $I_{\ell;i}$ and $J_{\ell;i}$ for $A_{\ell}$.


\IncMargin{1em}
\begin{algorithm}
\SetAlgoNoLine
\SetKwInOut{Output}{\textbf{Output}}

\BlankLine
Determine $\ell_{\max}$ and decompose the domain hierarchically\

Generate index sets $I_{1;ijk}$ and $J_{1;ijk}$\

$A_{1}\leftarrow A$

\For {$\ell = 1$ to $\ell_{\max}$ }{
     $A_{\ell+\frac{1}{2}} \leftarrow A_{\ell}(J_{\ell},J_{\ell})$\

     \For {($i,j,k$)$\in$ $\{$block index at level $\ell$$\}$}{
           $U_{\ell;ijk} \leftarrow A_{\ell}(I_{\ell;ijk},I_{\ell;ijk})$\

           $V_{\ell;ijk} \leftarrow A_{\ell}(J_{\ell;ijk},I_{\ell;ijk})$\

           Calculate $U_{\ell;ijk}^{-1}$\

           Calculate $K_{\ell;ijk}\leftarrow -V_{\ell;ijk}U_{\ell;ijk}^{-1}$\

           Calculate $A_{\ell+\frac{1}{2}}(J_{\ell;ijk},J_{\ell;ijk})\leftarrow A_{\ell+\frac{1}{2}}(J_{\ell;ijk},J_{\ell;ijk}) + K_{\ell;ijk}V_{\ell;ijk}^{T}$\
           }
          \If {$\ell<\ell_{\max}$}{
             Skeletonize cubic faces at level  $\ell+\frac{1}{2}$\

             \For {$s$ $\in$ $\{$block index at level $\ell+\frac{1}{2}$$\}$}{
                   Use ID to compute $T_{\ell+\frac{1}{2};s}$, $I_{\ell+\frac{1}{2};s}$ and $J_{\ell+\frac{1}{2};s}$

                   $U_{\ell+\frac{1}{2};s} \leftarrow A_{\ell+\frac{1}{2}}(I_{\ell+\frac{1}{2};s},I_{\ell+\frac{1}{2};s})$\

                   $V_{\ell+\frac{1}{2};s} \leftarrow A_{\ell+\frac{1}{2}}(J_{\ell+\frac{1}{2};s},I_{\ell+\frac{1}{2};s})$\

                   Calculate $\bar{\ell}_{\ell+\frac{1}{2};s}\leftarrow V_{\ell+\frac{1}{2};s}^{T}T_{\ell+\frac{1}{2};s}$\

                   Calculate $\bar{V}_{\ell+\frac{1}{2};s} \leftarrow V_{\ell+\frac{1}{2};s}-A_{\ell+\frac{1}{2}}(J_{\ell+\frac{1}{2};s},J_{\ell+\frac{1}{2};s})T_{\ell+\frac{1}{2};s}$\

                   Calculate $\bar{U}_{\ell+\frac{1}{2};s} \leftarrow U_{\ell+\frac{1}{2};s}- \bar{\ell}_{\ell+\frac{1}{2};s} - T_{\ell+\frac{1}{2};s}^{T}\bar{V}_{\ell+\frac{1}{2};s}$\
                   }
             $A_{\ell+1} \leftarrow A_{\ell+\frac{1}{2}}(J_{\ell+\frac{1}{2}},J_{\ell+\frac{1}{2}})$\

             \For {$s$ $\in$ $\{$block index at level $\ell+\frac{1}{2}$$\}$}{
                   Calculate $\bar{U}_{\ell+\frac{1}{2};s}^{-1}$\

                   Calculate $\bar{K}_{\ell+\frac{1}{2};s}\leftarrow -\bar{V}_{\ell+\frac{1}{2};s}\bar{U}_{\ell+\frac{1}{2};s}^{-1}$\

                   Calculate $A_{\ell+1}(J_{\ell+\frac{1}{2};s},J_{\ell+\frac{1}{2};s})\leftarrow A_{\ell+1}(J_{\ell+\frac{1}{2};s},J_{\ell+\frac{1}{2};s}) + \bar{K}_{\ell+\frac{1}{2};s}\bar{V}_{\ell+\frac{1}{2};s}^{T}$\
                  }
             Construct $I_{\ell+1}$ and $J_{\ell+1}$\
             }
     }
Calculate $G_{\ell_{\max}}\leftarrow A_{\ell_{\max}+\frac{1}{2}}^{-1}$

\Output {
        \\
        $I_{\ell},J_{\ell},I_{\ell+\frac{1}{2}},J_{\ell+\frac{1}{2}},U_{\ell;ijk}^{-1},\bar{U}_{\ell+\frac{1}{2};s}^{-1},K_{\ell;ijk},\bar{K}_{\ell+\frac{1}{2};s},G_{\ell_{\max}}$, for each $\ell,i,j,k,s$\\
}

\caption{Constructing the hierarchy of Schur complements of $A$ }
\label{algo1}
\end{algorithm}
\DecMargin{1em}




\subsection{Extracting the diagonals of the matrix inverse }
\label{ex}

After the hierarchy of Schur complements is constructed, one can proceed to extract the diagonals of the matrix $G$. It is important to note that computing the entire Schur complement $G_{\ell}$ is not required. 
This is based on the following observations,

\begin{equation}\numberwithin{equation}{section}
\label{obe}
\begin{cases}
G_{\ell-1}(I_{\ell;ijk}J_{\ell;ijk},I_{\ell;ijk}J_{\ell;ijk})\ \text{is\ determined\ by} \ G_{\ell-\frac{1}{2}}(J_{\ell;ijk},J_{\ell;ijk}),\\
G_{\ell-\frac{1}{2}}(I_{\ell-\frac{1}{2};i}J_{\ell-\frac{1}{2};i},I_{\ell-\frac{1}{2};i}J_{\ell-\frac{1}{2};i})\  \text{is\ determined\ by}\ G_{\ell}(J_{\ell-\frac{1}{2};i},J_{\ell-\frac{1}{2};i}).
\end{cases}
\end{equation}



For the purpose of extracting relevant information, we begin with considering the top level $\ell=L=3$.  $G_{5/2}$ can be calculated using the following formula with given $G_3$,

\begin{equation*}\numberwithin{equation}{section}
G_{\frac{5}{2}} = P_{3}
\left[\begin{array}{cc}
U_{3}^{-1}+U_{3}^{-1}V_{3}^{T}G_{3}V_{3}U_{3}^{-1}& -U_{3}^{-1}V_{3}^{T}G_{3}\\
-G_{3}V_{3}U_{3}^{-1} & G_{3}
\end{array}\right]P_{3}^{-1}.
\end{equation*}
The submatrices enclosed in the bracket are indexed by $(I_{3}|J_{3})$, while $G_{5/2}$ is indexed by $J_{5/2} = J_{5/2;1}J_{5/2;2}\dots J_{5/2;48}$, as a result of the permutation matrix $P_{3}$. However, it suffices to focus on $G_{5/2}(J_{5/2;i},J_{5/2;i})$ instead of the off-diagonal blocks to extract the diagonal elements of $G_{5/2}$. As a consequence, we can represent $G_{5/2}$ as:

\begin{equation*}\numberwithin{equation}{section}
G_{\frac{5}{2}} =
\left[\begin{array}{cccccc}
G_{\frac{5}{2};1} & * & * & * &* & * \\
* & G_{\frac{5}{2};2} & * & * &* &* \\
* & * & G_{\frac{5}{2};3} & * &* &*\\
* & * & * & G_{\frac{5}{2};4} &*&*\\
* & * & * & * &G_{\frac{5}{2};5}&* \\
* & * & * & * &*&G_{\frac{5}{2};48}
\end{array}\right]
\end{equation*}
with $G_{5/2;i} = G_{5/2}(J_{5/2;i},J_{5/2;i}), i=1,2,\dots,48.$

One recovers the elements of diagonal blocks for matrix $G_{5/2}$ in the previous layer. Furthermore, the  diagonal blocks of the following $G_2$ are acquired based on the observation of (\ref{obe}), 

\begin{equation}\numberwithin{equation}{section}
\label{eq:G2e}
G_{2} \approx P_{\frac{5}{2}}
\left[\begin{array}{cc}
\mathcal{H}_{2} & -\bar{U}_{\frac{5}{2}}^{-1}\bar{V}_{\frac{5}{2}}^{T}G_{\frac{5}{2}} + \mathcal{H}_{2}T_{\frac{5}{2}}^{T}\\
-G_{\frac{5}{2}}\bar{V}_{\frac{5}{2}}\bar{U}_{\frac{5}{2}}^{-1} + T_{\frac{5}{2}}\mathcal{H}_{2} & \mathfrak{H}_{2}
\end{array}\right]P_{\frac{5}{2}}^{-1}
\end{equation}
with

\begin{equation*}\numberwithin{equation}{section}
\mathcal{H}_{2}=\bar{U}_{5/2}^{-1}+\bar{U}_{5/2}^{-1}\bar{V}_{5/2}^{T}G_{5/2}\bar{V}_{5/2}\bar{U}_{5/2}^{-1},
\end{equation*}
and

\begin{equation*}\numberwithin{equation}{section}
\mathfrak{H}_{2} = T_{\frac{5}{2}}\mathcal{H}_{2}T_{\frac{5}{2}}^{T}-G_{\frac{5}{2}}\bar{V}_{\frac{5}{2}}\bar{U}_{\frac{5}{2}}^{-1}T_{\frac{5}{2}}^{T}-T_{\frac{5}{2}}\bar{U}_{\frac{5}{2}}^{-1}\bar{V}_{\frac{5}{2}}^{T}G_{\frac{5}{2}}+G_{\frac{5}{2}}.
\end{equation*}
All matrices in the bracket of (\ref{eq:G2e}) are indexed by $(I_{5/2}|J_{5/2})$. $G_2$ is indexed by 

\begin{equation*}\numberwithin{equation}{section}
J_{2} = J_{2;111}J_{2;121}J_{2;211}J_{2;221}J_{2;112}J_{2;122}J_{2;212}J_{2;222}
\end{equation*}
due to the permutation matrix $P_{5/2}$. \\
Recalling the construction process, one can assert that $T_{5/2}$, $U_{5/2}^{-1}$, and $V_{5/2}$ are block diagonal matrices and the diagonal blocks of the following matrices are both obtained,

\begin{equation*}\numberwithin{equation}{section}
\begin{split}
\bar{U}_{\frac{5}{2}}^{-1}\bar{V}_{\frac{5}{2}}^{T}G_{\frac{5}{2}}\bar{V}_{\frac{5}{2}}\bar{U}_{\frac{5}{2}}^{-1} =
\left[\begin{array}{ccc}
\bar{U}_{\frac{5}{2};1}^{-1}\bar{V}_{\frac{5}{2};1}^{T}G_{\frac{5}{2};1}\bar{V}_{\frac{5}{2};1}\bar{U}_{\frac{5}{2};1}^{-1}& \cdots & *\\
\vdots & \ddots & \vdots\\
* & \cdots & \bar{U}_{\frac{5}{2};48}^{-1}\bar{V}_{\frac{5}{2};48}^{T}G_{\frac{5}{2};48}\bar{V}_{\frac{5}{2};48}\bar{U}_{\frac{5}{2};48}^{-1}
\end{array}\right],
\end{split}
\end{equation*} 
\begin{equation*}\numberwithin{equation}{section}
\begin{split}
G_{\frac{5}{2}}\bar{V}_{\frac{5}{2}}\bar{U}_{\frac{5}{2}}^{-1}T_{\frac{5}{2}}^{T} =
\left[\begin{array}{ccc}
G_{\frac{5}{2};1}\bar{V}_{\frac{5}{2};1}\bar{U}_{\frac{5}{2};1}^{-1}T_{\frac{5}{2};1}^{T}& \cdots & *\\
\vdots & \ddots & \vdots\\
* & \cdots & G_{\frac{5}{2};48}\bar{V}_{\frac{5}{2};48}\bar{U}_{\frac{5}{2};48}^{-1}T_{\frac{5}{2};48}^{T}
\end{array}\right].
\end{split}
\end{equation*}
This means that only block-block multiplication is needed to get the elements of the diagonal blocks $\mathcal{H}_{2}$ and $\mathfrak{H}_{2}$. The computational complexity is then greatly reduced. Additionally, the diagonal blocks $G_{2}(J_{2;ijk},J_{2;ijk})$ are obtained directly.

\IncMargin{1em}
\begin{algorithm}
\SetAlgoNoLine
\SetKwInOut{Input}{\textbf{Input}}

\BlankLine

\Input{
        \\
        Output of Algorithm \ref{algo1}\\}
\For {$\ell = \ell_{\max}$ to $1$}{
      \For {($i,j,k$)$\in$ $\{$block index at level $\ell$$\}$}{
            Calculate $G_{\ell-\frac{1}{2}}(I_{\ell;ijk},I_{\ell;ijk})\leftarrow U_{\ell;ijk}^{-1}+K_{\ell;ijk}^{T}G_{\ell}(J_{\ell;ijk},J_{\ell;ijk})K_{\ell;ijk}$\

            Calculate $G_{\ell-\frac{1}{2}}(J_{\ell;ijk},I_{\ell;ijk})\leftarrow G_{\ell}(J_{\ell;ijk},J_{\ell;ijk})K_{\ell;ijk}$\

            $G_{\ell-\frac{1}{2}}(I_{\ell;ijk},J_{\ell;ijk})\leftarrow G_{\ell-\frac{1}{2}}(J_{\ell;ijk},I_{\ell;ijk})^{T}$\

            $G_{\ell-\frac{1}{2}}(J_{\ell;ijk},J_{\ell;ijk})\leftarrow G_{\ell}(J_{\ell;ijk},J_{\ell;ijk})$

      }

      \If {$\ell>1$}{
          \For {$s$ $\in$ $\{$block index at level $\ell-\frac{1}{2}$$\}$}{
                Calculate $G_{\ell-1}(I_{\ell-\frac{1}{2};s},I_{\ell-\frac{1}{2};s})\leftarrow \bar{U}_{\ell-\frac{1}{2};s}^{-1}+\bar{K}_{\ell-\frac{1}{2};s}^{T}G_{\ell-\frac{1}{2}}(J_{\ell-\frac{1}{2};s},J_{\ell-\frac{1}{2};s})\bar{K}_{\ell-\frac{1}{2};s}$\

                Calculate $\bar{W}_{\ell-1}(J_{\ell-\frac{1}{2};s},I_{\ell-\frac{1}{2};s})\leftarrow G_{\ell-\frac{1}{2}}\bar{K}_{\ell-\frac{1}{2};s}$\

                $G_{\ell-1}(J_{\ell-\frac{1}{2};s},I_{\ell-\frac{1}{2};s})\leftarrow \bar{W}_{\ell-1}(J_{\ell-\frac{1}{2};s},I_{\ell-\frac{1}{2};s})+T_{\ell-\frac{1}{2};s}G_{\ell-1}(I_{\ell-\frac{1}{2};s},I_{\ell-\frac{1}{2};s})$\

                $G_{\ell-1}(I_{\ell-\frac{1}{2};s},J_{\ell-\frac{1}{2};s})\leftarrow G_{\ell-1}(J_{\ell-\frac{1}{2};s},I_{\ell-\frac{1}{2};s})^{T}$\

                $G_{\ell-1}(J_{\ell-\frac{1}{2};s},J_{\ell-\frac{1}{2};s})\leftarrow G_{\ell-1}(J_{\ell-\frac{1}{2};s},I_{\ell-\frac{1}{2};s})T_{\ell-\frac{1}{2};s}^{T}+T_{\ell-\frac{1}{2};s}\bar{W}_{\ell-1}(J_{\ell-\frac{1}{2};s},I_{\ell-\frac{1}{2};s})+G_{\ell-\frac{1}{2}}(J_{\ell-\frac{1}{2};s},J_{\ell-\frac{1}{2};s})$
          }
      }
}
\caption{Extracting the diagonals of $A^{-1}$}
\label{algo2}
\end{algorithm}
\DecMargin{1em}

At Level 2, one has

\begin{equation}\numberwithin{equation}{section}\label{eq:G3/2}
G_{\frac{3}{2}} = P_{2}
\left[\begin{array}{cc}
U_{2}^{-1}+U_{2}^{-1}V_{2}^{T}G_{2}V_{2}U_{2}^{-1}& -U_{2}^{-1}V_{2}^{T}G_{2}\\
-G_{2}V_{2}U_{2}^{-1} & G_{2}
\end{array}\right]P_{2}^{-1}.
\end{equation}
Similarly, the submatrices in the bracket of (\ref{eq:G3/2}) are indexed by $(I_{2}|J_{2})$.  $G_{3/2}$ is indexed by $J_{3/2} = J_{3/2;1}\cdots J_{3/2;384}$ with the permutation matrix $P_2$. Moreover, one just needs to compute $G_{3/2}(J_{3/2;i},J_{3/2;i})$ in this step.\\


At Level $3/2$, one has

\begin{equation}\numberwithin{equation}{section}
\label{eq:G1e}
G_{1} \approx P_{\frac{3}{2}}
\left[\begin{array}{cc}
\mathcal{H}_{1} & -\bar{U}_{\frac{3}{2}}^{-1}\bar{V}_{\frac{3}{2}}^{T}G_{\frac{3}{2}} + \mathcal{H}_{1}T_{\frac{3}{2}}^{T}\\
-G_{\frac{3}{2}}\bar{V}_{\frac{3}{2}}\bar{U}_{\frac{3}{2}}^{-1} + T_{\frac{3}{2}}\mathcal{H}_{1} & \mathfrak{H}_{1}
\end{array}\right]P_{\frac{3}{2}}^{-1},
\end{equation}
with

\begin{equation*}\numberwithin{equation}{section}
\mathcal{H}_{1}=\bar{U}_{\frac{3}{2}}^{-1}+\bar{U}_{\frac{3}{2}}^{-1}\bar{V}_{\frac{3}{2}}^{T}G_{\frac{3}{2}}\bar{V}_{\frac{3}{2}}\bar{U}_{\frac{3}{2}}^{-1},
\end{equation*}
and

\begin{equation*}\numberwithin{equation}{section}
\mathfrak{H}_{1} = T_{\frac{3}{2}}\mathcal{H}_{1}T_{\frac{3}{2}}^{T}-G_{\frac{3}{2}}\bar{V}_{\frac{3}{2}}\bar{U}_{\frac{3}{2}}^{-1}T_{\frac{3}{2}}^{T}-T_{\frac{3}{2}}\bar{U}_{\frac{3}{2}}^{-1}\bar{V}_{\frac{3}{2}}^{T}G_{\frac{3}{2}}+G_{\frac{3}{2}}.
\end{equation*}
The diagonal blocks of $\mathcal{H}_{1}$ and $\mathfrak{H}_{1}$ can be efficiently computed using block-block multiplication, just like the Level $\ell=5/2$. The index of the submatrices in the bracket of equation (\ref{eq:G1e}) is  
$(I_{3/2}|J_{3/2})$. Due to the permutation matrix $P_{3/2}$, matrix $G_1$ is indexed by $J_{1} = J_{1;111}J_{1;121}\cdots J_{1;444}$ and all elements needed are the diagonal blocks $G_{1}(J_{1;ijk},J_{1;ijk})$.

At the bottom level $\ell=1$, one applies the same procedure as at Level 2 and Level 3. Specifically, one obtains $G_{1}(J_{1;ijk},J_{1;ijk})$ from Level $3/2$, while $G(J_{0;ijk},J_{0;ijk})$ is computed directly. Consequently, the diagonal elements of $G$ can be obtained by combining the diagonal elements of each level.\\
Finally, a quasilinear scaling algorithm can be achieved to extract the diagonal elements of $G$ recursively. Algorithm \ref{algo2} presents the organized form of this procedure. It is worth noting that the reindexing process is implicitly included in Algorithm \ref{algo2} as one uses the index sets $J_{\ell;ijk}$ or $J_{\ell;i}$ for $G_{\ell}$.

\subsection{The SelInvHIF with edge skeletonization}
\label{SHIF-edge}

In Section \ref{21}, the construction step of the SelInvHIF is characterized by its incorporation of interior points elimination and face skeletonization, while SelInvHIF with edge skeletonization further advances this approach by implementing additional edge skeletonization.  
That is, additional layers are also introduced to skeletonize edges to achieve a complete dimensionality reduction. 
Let us recall the construction step within the SelInvHIF, wherein the hierarchy construction of Schur complements is systematically performed at levels $1$, $3/2$, $2$, $5/2$, $\dots$, and $L$. 
Compared to SelInvHIF, the construction step of SelInvHIF with edge skeletonization is carried out at levels $1$, $4/3$, $5/3$,$2$, $7/3$, $\dots$, and $L$.
Specifically, At each integer level, the points are reindexed and the interior points are eliminated accordingly.
In addition, the face skeletonization is performed at level $\ell+1/3$ and edge skeletonization is performed 
at level $\ell+2/3$, respectively. Precisely, the relationships between levels are defined the same as \ref{eq:rel}. Furthermore, one obtains the following recursive relation with integer $\ell$,

\begin{equation*}\numberwithin{equation}{section}
G_{\ell-1} \approx P_{\ell-\frac{2}{3}}Q_{\ell-\frac{2}{3}}L_{\ell-\frac{2}{3}}^{T}
\left[\begin{array}{cc}
\bar{U}_{\ell-\frac{2}{3}}^{-1} &  \\
 & G_{\ell-\frac{2}{3}}
\end{array}\right]L_{\ell-\frac{2}{3}}Q_{\ell-\frac{2}{3}}^{T}P_{\ell-\frac{2}{3}}^{-1},
\end{equation*}

\begin{equation*}\numberwithin{equation}{section}
G_{\ell-\frac{2}{3}} \approx P_{\ell-\frac{1}{3}}Q_{\ell-\frac{1}{3}}L_{\ell-\frac{1}{3}}^{T}
\left[\begin{array}{cc}
\bar{U}_{\ell-\frac{1}{3}}^{-1} &  \\
 & G_{\ell-\frac{1}{3}}
\end{array}\right]L_{\ell-\frac{1}{3}}Q_{\ell-\frac{1}{3}}^{T}P_{\ell-\frac{1}{3}}^{-1},
\end{equation*}

\begin{equation*}\numberwithin{equation}{section}
G_{\ell-\frac{1}{3}} = P_{\ell}L_{\ell}^{T}
\left[\begin{array}{cc}
U_{\ell}^{-1} &  \\
 & G_{\ell}
\end{array}\right]L_{\ell}P_{\ell}^{-1}.
\end{equation*}

Similar to Section \ref{ex}, one can extract the diagonals of the matrix $G$ based on the hierarchy of Schur complements. The following observations shows that computing the entire $G_{\ell}$ is not required,

\begin{equation}\numberwithin{equation}{section}
\label{obe2}
\begin{cases}
G_{\ell-1}(I_{\ell;ijk}J_{\ell;ijk},I_{\ell;ijk}J_{\ell;ijk})\ \text{is\ determined\ by} \ G_{\ell-\frac{2}{3}}(J_{\ell;ijk},J_{\ell;ijk}),\\
G_{\ell-\frac{2}{3}}(I_{\ell-\frac{2}{3};i}J_{\ell-\frac{2}{3};i},I_{\ell-\frac{2}{3};i}J_{\ell-\frac{2}{3};i})\  \text{is\ determined\ by}\ G_{\ell-\frac{1}{3}}(J_{\ell-\frac{2}{3};i},J_{\ell-\frac{2}{3};i}),\\
G_{\ell-\frac{1}{3}}(I_{\ell-\frac{1}{3};i}J_{\ell-\frac{1}{3};i},I_{\ell-\frac{1}{3};i}J_{\ell-\frac{1}{3};i})\  \text{is\ determined\ by}\ G_{\ell}(J_{\ell-\frac{1}{3};i},J_{\ell-\frac{1}{3};i}).
\end{cases}
\end{equation}
Based on \ref{obe2}, the recovery process for $G$ can be carried out as in Section \ref{ex}. Hence, the construction step and extracting step of the SelInvHIF with edge skeletonization can be described in Algorithms \ref{algo1-edge} and \ref{algo2-edge}. 

\IncMargin{1em}
\begin{algorithm}
\SetAlgoNoLine
\SetKwInOut{Output}{\textbf{Output}}

\BlankLine
Determine $\ell_{\max}$, the hierarchical structure of domain, and index sets $I_{1;ijk}$ and $J_{1;ijk}$\

\For {$\ell = 1$ to $\ell_{\max}$ }{
     $A_{\ell+\frac{1}{3}} \leftarrow A_{\ell}(J_{\ell},J_{\ell})$\

     \For {($i,j,k$)$\in$ $\{$block index at level $\ell$$\}$}{
           $U_{\ell;ijk} \leftarrow A_{\ell}(I_{\ell;ijk},I_{\ell;ijk})$, $V_{\ell;ijk} \leftarrow A_{\ell}(J_{\ell;ijk},I_{\ell;ijk})$\
           
           Calculate $U_{\ell;ijk}^{-1}$, $K_{\ell;ijk}\leftarrow -V_{\ell;ijk}U_{\ell;ijk}^{-1}$\
           
           Calculate $A_{\ell+\frac{1}{3}}(J_{\ell;ijk},J_{\ell;ijk})\leftarrow A_{\ell+\frac{1}{3}}(J_{\ell;ijk},J_{\ell;ijk}) + K_{\ell;ijk}V_{\ell;ijk}^{T}$\
           }
          \If {$\ell<\ell_{\max}$}{
             Skeletonize cubic faces at level  $\ell+\frac{1}{3}$\

             \For {$s$ $\in$ $\{$block index at level $\ell+\frac{1}{3}$$\}$}{
                   Use ID to compute $T_{\ell+\frac{1}{3};s}$, $I_{\ell+\frac{1}{3};s}$ and $J_{\ell+\frac{1}{3};s}$

                   $U_{\ell+\frac{1}{3};s} \leftarrow A_{\ell+\frac{1}{3}}(I_{\ell+\frac{1}{3};s},I_{\ell+\frac{1}{3};s})$,
                   $V_{\ell+\frac{1}{3};s} \leftarrow A_{\ell+\frac{1}{3}}(J_{\ell+\frac{1}{3};s},I_{\ell+\frac{1}{3};s})$\

                   Calculate $\bar{\ell}_{\ell+\frac{1}{3};s}\leftarrow V_{\ell+\frac{1}{3};s}^{T}T_{\ell+\frac{1}{3};s}$\

                   Calculate $\bar{V}_{\ell+\frac{1}{3};s} \leftarrow V_{\ell+\frac{1}{3};s}-A_{\ell+\frac{1}{3}}(J_{\ell+\frac{1}{3};s},J_{\ell+\frac{1}{3};s})T_{\ell+\frac{1}{3};s}$\

                   Calculate $\bar{U}_{\ell+\frac{1}{3};s} \leftarrow U_{\ell+\frac{1}{3};s}- \bar{\ell}_{\ell+\frac{1}{3};s} - T_{\ell+\frac{1}{3};s}^{T}\bar{V}_{\ell+\frac{1}{3};s}$\
                   }
             $A_{\ell+\frac{2}{3}} \leftarrow A_{\ell+\frac{1}{3}}(J_{\ell+\frac{1}{3}},J_{\ell+\frac{1}{3}})$\

             \For {$s$ $\in$ $\{$block index at level $\ell+\frac{1}{3}$$\}$}{
                   Calculate $\bar{U}_{\ell+\frac{1}{3};s}^{-1}$,
                   $\bar{K}_{\ell+\frac{1}{3};s}\leftarrow -\bar{V}_{\ell+\frac{1}{3};s}\bar{U}_{\ell+\frac{1}{3};s}^{-1}$\

                   Calculate $A_{\ell+1}(J_{\ell+\frac{1}{3};s},J_{\ell+\frac{1}{3};s})\leftarrow A_{\ell+1}(J_{\ell+\frac{1}{3};s},J_{\ell+\frac{1}{3};s}) + \bar{K}_{\ell+\frac{1}{3};s}\bar{V}_{\ell+\frac{1}{3};s}^{T}$\
                  }
             Construct $I_{\ell+\frac{2}{3}}$ and $J_{\ell+\frac{2}{3}}$\
             }
                       \If {$\ell<\ell_{\max}$}{
             Skeletonize cubic edges at level  $\ell+\frac{2}{3}$\

             \For {$s$ $\in$ $\{$block index at level $\ell+\frac{2}{3}$$\}$}{
                   Use ID to compute $T_{\ell+\frac{2}{3};s}$, $I_{\ell+\frac{2}{3};s}$ and $J_{\ell+\frac{2}{3};s}$

                   $U_{\ell+\frac{2}{3};s} \leftarrow A_{\ell+\frac{2}{3}}(I_{\ell+\frac{2}{3};s},I_{\ell+\frac{2}{3};s})$,
                   $V_{\ell+\frac{2}{3};s} \leftarrow A_{\ell+\frac{2}{3}}(J_{\ell+\frac{2}{3};s},I_{\ell+\frac{2}{3};s})$\

                   Calculate $\bar{\ell}_{\ell+\frac{2}{3};s}\leftarrow V_{\ell+\frac{2}{3};s}^{T}T_{\ell+\frac{2}{3};s}$\

                   Calculate $\bar{V}_{\ell+\frac{2}{3};s} \leftarrow V_{\ell+\frac{2}{3};s}-A_{\ell+\frac{2}{3}}(J_{\ell+\frac{2}{3};s},J_{\ell+\frac{2}{3};s})T_{\ell+\frac{2}{3};s}$\

                   Calculate $\bar{U}_{\ell+\frac{2}{3};s} \leftarrow U_{\ell+\frac{2}{3};s}- \bar{\ell}_{\ell+\frac{2}{3};s} - T_{\ell+\frac{2}{3};s}^{T}\bar{V}_{\ell+\frac{2}{3};s}$\
                   }
             $A_{\ell+1} \leftarrow A_{\ell+\frac{2}{3}}(J_{\ell+\frac{2}{3}},J_{\ell+\frac{2}{3}})$\

             \For {$s$ $\in$ $\{$block index at level $\ell+\frac{2}{3}$$\}$}{
                   Calculate $\bar{U}_{\ell+\frac{2}{3};s}^{-1}$,
                   $\bar{K}_{\ell+\frac{2}{3};s}\leftarrow -\bar{V}_{\ell+\frac{2}{3};s}\bar{U}_{\ell+\frac{2}{3};s}^{-1}$\

                   Calculate $A_{\ell+1}(J_{\ell+\frac{2}{3};s},J_{\ell+\frac{2}{3};s})\leftarrow A_{\ell+1}(J_{\ell+\frac{2}{3};s},J_{\ell+\frac{2}{3};s}) + \bar{K}_{\ell+\frac{2}{3};s}\bar{V}_{\ell+\frac{2}{3};s}^{T}$\
                  }
             Construct $I_{\ell+1}$ and $J_{\ell+1}$\
             }
     }

\Output {
        \\
        $I_{\ell},J_{\ell},I_{\ell+\frac{1}{3}},J_{\ell+\frac{1}{3}}, I_{\ell+\frac{2}{3}},J_{\ell+\frac{2}{3}}, U_{\ell;ijk}^{-1},\bar{U}_{\ell+\frac{1}{3};s}^{-1}, \bar{U}_{\ell+\frac{2}{3};s}^{-1}, K_{\ell;ijk},\bar{K}_{\ell+\frac{1}{3};s},  \bar{K}_{\ell+\frac{2}{3};s}^{-1}, G_{\ell_{\max}}$
}

\caption{The hierarchy of Schur complements of $A$ with edge skeletonization}
\label{algo1-edge}
\end{algorithm}
\DecMargin{1em}

\IncMargin{1em}
\begin{algorithm}
\SetAlgoNoLine
\SetKwInOut{Input}{\textbf{Input}}

\BlankLine

\Input{
        \\
        Output of Algorithm \ref{algo1-edge}\\}
\For {$\ell = \ell_{\max}$ to $1$}{
      \For {($i,j,k$)$\in$ $\{$block index at level $\ell$$\}$}{
            Calculate $G_{\ell-\frac{1}{3}}(I_{\ell;ijk},I_{\ell;ijk})\leftarrow U_{\ell;ijk}^{-1}+K_{\ell;ijk}^{T}G_{\ell}(J_{\ell;ijk},J_{\ell;ijk})K_{\ell;ijk}$\

            Calculate $G_{\ell-\frac{1}{3}}(J_{\ell;ijk},I_{\ell;ijk})\leftarrow G_{\ell}(J_{\ell;ijk},J_{\ell;ijk})K_{\ell;ijk}$\

            $G_{\ell-\frac{1}{3}}(I_{\ell;ijk},J_{\ell;ijk})\leftarrow G_{\ell-\frac{1}{3}}(J_{\ell;ijk},I_{\ell;ijk})^{T}$\

            $G_{\ell-\frac{1}{3}}(J_{\ell;ijk},J_{\ell;ijk})\leftarrow G_{\ell}(J_{\ell;ijk},J_{\ell;ijk})$

      }

      \If {$\ell>1$}{
          \For {$s$ $\in$ $\{$block index at level $\ell-\frac{1}{3}$$\}$}{
                Calculate $G_{\ell-\frac{2}{3}}(I_{\ell-\frac{1}{3};s},I_{\ell-\frac{1}{3};s})\leftarrow \bar{U}_{\ell-\frac{1}{3};s}^{-1}+\bar{K}_{\ell-\frac{1}{3};s}^{T}G_{\ell-\frac{1}{3}}(J_{\ell-\frac{1}{3};s},J_{\ell-\frac{1}{3};s})\bar{K}_{\ell-\frac{1}{3};s}$\

                Calculate $\bar{W}_{\ell-\frac{2}{3}}(J_{\ell-\frac{1}{3};s},I_{\ell-\frac{1}{3};s})\leftarrow G_{\ell-\frac{1}{3}}\bar{K}_{\ell-\frac{1}{3};s}$\

                $G_{\ell-\frac{2}{3}}(J_{\ell-\frac{1}{3};s},I_{\ell-\frac{1}{3};s})\leftarrow \bar{W}_{\ell-1}(J_{\ell-\frac{1}{3};s},I_{\ell-\frac{1}{3};s})+T_{\ell-\frac{1}{3};s}G_{\ell-\frac{2}{3}}(I_{\ell-\frac{1}{3};s},I_{\ell-\frac{1}{3};s})$\

                $G_{\ell-\frac{2}{3}}(I_{\ell-\frac{1}{3};s},J_{\ell-\frac{1}{3};s})\leftarrow G_{\ell-\frac{2}{3}}(J_{\ell-\frac{1}{3};s},I_{\ell-\frac{1}{3};s})^{T}$\

                $G_{\ell-\frac{2}{3}}(J_{\ell-\frac{1}{3};s},J_{\ell-\frac{1}{3};s})\leftarrow G_{\ell-\frac{2}{3}}(J_{\ell-\frac{1}{3};s},I_{\ell-\frac{1}{3};s})T_{\ell-\frac{1}{3};s}^{T}+T_{\ell-\frac{1}{3};s}\bar{W}_{\ell-1}(J_{\ell-\frac{1}{3};s},I_{\ell-\frac{1}{3};s})+G_{\ell-\frac{1}{3}}(J_{\ell-\frac{1}{3};s},J_{\ell-\frac{1}{3};s})$
          }
      }
   \If {$\ell>1$}{
          \For {$s$ $\in$ $\{$block index at level $\ell-\frac{2}{3}$$\}$}{
                Calculate $G_{\ell-1}(I_{\ell-\frac{2}{3};s},I_{\ell-\frac{2}{3};s})\leftarrow \bar{U}_{\ell-\frac{2}{3};s}^{-1}+\bar{K}_{\ell-\frac{2}{3};s}^{T}G_{\ell-\frac{2}{3}}(J_{\ell-\frac{2}{3};s},J_{\ell-\frac{2}{3};s})\bar{K}_{\ell-\frac{2}{3};s}$\

                Calculate $\bar{W}_{\ell-1}(J_{\ell-\frac{2}{3};s},I_{\ell-\frac{2}{3};s})\leftarrow G_{\ell-\frac{2}{3}}\bar{K}_{\ell-\frac{2}{3};s}$\

                $G_{\ell-1}(J_{\ell-\frac{2}{3};s},I_{\ell-\frac{2}{3};s})\leftarrow \bar{W}_{\ell-1}(J_{\ell-\frac{2}{3};s},I_{\ell-\frac{2}{3};s})+T_{\ell-\frac{2}{3};s}G_{\ell-1}(I_{\ell-\frac{2}{3};s},I_{\ell-\frac{2}{3};s})$\

                $G_{\ell-1}(I_{\ell-\frac{2}{3};s},J_{\ell-\frac{2}{3};s})\leftarrow G_{\ell-1}(J_{\ell-\frac{2}{3};s},I_{\ell-\frac{2}{3};s})^{T}$\

                $G_{\ell-1}(J_{\ell-\frac{2}{3};s},J_{\ell-\frac{2}{3};s})\leftarrow G_{\ell-1}(J_{\ell-\frac{2}{3};s},I_{\ell-\frac{2}{3};s})T_{\ell-\frac{2}{3};s}^{T}+T_{\ell-\frac{2}{3};s}\bar{W}_{\ell-1}(J_{\ell-\frac{2}{3};s},I_{\ell-\frac{2}{3};s})+G_{\ell-\frac{2}{3}}(J_{\ell-\frac{2}{3};s},J_{\ell-\frac{2}{3};s})$
          }
      }
}
\caption{Extracting the diagonals of $A^{-1}$ with edge skeletonization}
\label{algo2-edge}
\end{algorithm}
\DecMargin{1em}

\subsection{Computational Complexity}

In this section, the computational complexity of the SelInvHIF is considered. Assume that the domain consists of  $N = \sqrt[3]{N}\times \sqrt[3]{N}\times \sqrt[3]{N}$ points and  set $\sqrt[3]{N}=2^{L}$ with $\ell_{\max}<L$. The number of blocks at level $\ell$ is defined as $n_{B}(\ell)$, and the following formula holds

\begin{equation*}\numberwithin{equation}{section}
n_{B}(\ell) =O(2^{3(\ell_{\max}-\ell)}).
\end{equation*}

The number of points in each block (cubic face or cubic edge) is represented as $n_{P}(\ell).$ It should be noted that the interior or redundant points from the previous level are not included, since they have already been eliminated in previous levels.
In order to estimate $n_{P}(\ell)$, we rely on the assumption made in \cite{hifde} regarding the skeletonization. 
The  assumption states that the typical skeleton size is:


\begin{equation*}\numberwithin{equation}{section}
n_{P}(\ell) =
\begin{cases}
O(\ell),& \text {for edges;} \\
O(2^\ell),& \text {for faces.}
\end{cases}
\end{equation*}

Firstly, we consider the construction step of SelInvHIF, which involves the following steps in Algorithm \ref{algo1}. 
At the integer level $\ell$, one computes $U_{\ell;ijk}^{-1}$ (Line 9) for each block.
One then multiplies the inverse with $V_{\ell;ijk}$ to obtain $K_{\ell;ijk}$ (Line 10) and update the new $A_{\ell+1/2}(J_{\ell;ijk},J_{\ell;ijk})$ (Line 11). 
At fractional level $\ell+1/2$, the $T_{\ell+1/2;k}$ is recorded by ID for each cell (Line 16). 
The cost for this step is $O(n_{P}(\ell)^{3})$ since each cell only interacts with $O(1)$ cells, one then applies it (Lines 19, 20, and 21) and multiply the inverse of $\bar{U}_{\ell+1/2;k}$ (Line 25) with $\bar{V}_{\ell+1/2;k}$ to obtain $\bar{K}_{\ell+1/2;k}$ (Line 26).
Finally, one update $A_{\ell+1}(J_{\ell+1/2;k},J_{\ell+1/2;k})$ (Line 27). Thus, the computational cost for these steps at each level is $O(n_{P}(\ell)^{3})$. The total computational complexity is 
 
\begin{equation*}\numberwithin{equation}{section}\label{ineq:complexity}
\begin{split}
\sum\limits_{\ell=1}^{\ell_{\max}}n_{B}(\ell)n_{P}(\ell)^{3} \leq C\sum\limits_{\ell=1}^{\ell_{\max}}2^{3 \ell_{\max}-3\ell}2^{3 \ell}\leq C_0 N\log N,
\end{split}
\end{equation*}
where $C$ and $C_0$ are constant. The total computational cost for the construction step is $O(N\log N)$ with $\ell_{\max}=O(L)$.

In addition, the extraction phase is considered and the following steps are shown in  Algorithm \ref{algo2}. At the integer level $\ell$, one can calculate $G_{\ell-1/2}(I_{\ell;ijk},I_{\ell;ijk})$ (Line $3$) and $G_{\ell- 1/2}(J_{\ell;ijk},I_{\ell;ijk})$ (Line $4$) for each block. 
At the fractional level $\ell- 1/2$,  $G_{\ell-1}(I_{\ell-1/2;k},I_{\ell-1/2;k})$ (Line $10$), \\
$G_{\ell-1}(J_{\ell-\frac{1}{2};k},I_{\ell-1/2;k})$ (Line $12$) and $G_{\ell-1}(J_{\ell-1/2;k},J_{\ell-1/2;k})$ (Line $14$) are calculated  for each cell. The computational cost for these steps at each level is $O(n_{P}(\ell)^{3})$.  It turns out that the complexity for the extraction phase is also $O(N\log N)$.

As for the computational complexity of SelInvHIF with edge skeletonization, the cost for each step at each level is also $n_{B}(\ell)n_{P}(\ell)^{3}$. Thus, the total computational complexity is 
 
\begin{equation*}\numberwithin{equation}{section}\label{ineq:complexity}
\begin{split}
\sum\limits_{\ell=1}^{\ell_{\max}}n_{B}(\ell)n_{P}(\ell)^{3} \leq C\sum\limits_{\ell=1}^{\ell_{\max}}2^{3 \ell_{\max}-3\ell} \ell^3\leq C_0 N,
\end{split}
\end{equation*}
where $C$ and $C_0$ are constant. The total computational cost for the construction step is $O(N)$ with $\ell_{\max}=O(L)$.

Finally, the quasi-linear scaling of  the SelInvHIF and the linear scaling of the SelInvHIF with edge skeletonization are proved. Although SelInvHIF with edge skeletonization can achieve $O(N)$ complexity, some fill-in is generated after edge  skeletonization, which brings additional computational costs. This also implies that it can achieve optimal complexity only if $N$ is suitably large, which presents challenges in directly applying it to the MPB equation of interest. Therefore, only the SelInvHIF is applied in all subsequent numerical examples.


\section{Numerical Method for MPB Equations}
\label{NMmPBE}

In this section, the iterative solver is proposed to solve the MPB equation. The following governing equations for the whole space in \cite{pb1} based on Gaussian variational field theory \cite{pnp1,Podgornik1989ElectrostaticCF},

\begin{equation}
\label{eq:ite}
\left\{\begin{array}{l}
{\nabla \cdot \eta(\bm{r}) \nabla \phi-\chi \Lambda e^{-\Xi c(\bm{r}) / 2} \sinh \phi=-2 \rho_{f}(\bm{r})}, \\
{\left[\nabla \cdot \eta(\bm{r}) \nabla-\chi \Lambda e^{-\Xi c(\bm{r}) / 2} \cosh \phi\right] G\left(\bm{r}, \bm{r}^{\prime}\right)=-4 \pi \delta\left(\bm{r}-\bm{r}^{\prime}\right)}, \\
\end{array}\right.
\end{equation}
with the potential $\phi$, the relative dielectric function  $\eta(\bm{r})$, the density of fixed charge $ \rho_f(\bm{r})$,  and the Green's function $G\left(\bm{r}, \bm{r}^{\prime}\right)$. The coupling parameter $\Xi$ and the rescaled fugacity $ \Lambda$ are given for specific problems. 
The function $\chi(\bm{r})$ is defined as 1 to represent the region that is accessible for ions, while it is defined as 0 elsewhere.
The correlation function ${c(\bm{r})}$  reads

\begin{equation*}
{c(\bm{r})=\lim _{\bm{r}\rightarrow \bm{r}^{\prime}}\left[G\left(\bm{r}, \bm{r}^{\prime}\right)-1 /\eta(\bm{r})\left|\bm{r}-\bm{r}^{\prime}\right| \right]}.
\end{equation*}

A self-consistent iterative scheme is utilized to solve the partial differential equations (\ref{eq:ite}), as described in \cite{pb1}. This scheme comprises of two alternating steps: first, given a ${c(\bm{r})}$ , the modified Poisson-Boltzmann (PB) equation (the first equation) is solved to obtain the potential $\phi$ with given boundary conditions. Second, for a given ${c(\bm{r})}$ and $\phi$, the GDH equation (the second equation) is solved to obtain $G$ and a new ${c(\bm{r})}$. These two steps are iterated until the solution reaches the desired convergence criteria. The iterative scheme is mathematically expressed \cite{pnp1,pnp2}:

\begin{equation}
\left\{\begin{array}{l}{\nabla \cdot \eta(\bm{r}) \nabla \phi^{(k+1)}-\Lambda e^{-\frac{\Xi c^{(k)}}{2}} \sinh \phi^{(k+1)}=-2 \rho_{f}(\bm{r})},\\
{\left[\nabla \cdot \eta(\bm{r}) \nabla-\Lambda e^{-\frac{\Xi c^{(k)}}{2}} \cosh \phi^{(k+1)}\right] G^{(k+1)}=-4 \pi \delta\left(\bm{r}-\bm{r}^{\prime}\right)},\\
{c^{(k+1)}(\bm{r})=\lim _{\bm{r}\rightarrow \bm{r}^{\prime}}\left[G^{(k+1)}\left(\bm{r}, \bm{r}^{\prime}\right)-1 /\eta(\bm{r})\left|\bm{r}-\bm{r}^{\prime}\right|\right]},
\end{array}\right.
\end{equation}
where the superscript $k=0,1,...,K$ indicates the $k$th iteration step.

In addition, the PB steps can be efficiently solved using standard direct solvers. However, the key of solving a self-consistent equation lies in the GDH equation, which can be reformulated during a self-consistent iteration. In three dimensions, a finite difference approximation of the equation is employed to obtain the following algebraic equation:
\begin{equation}
\bm{A G}=\bm{I}
\end{equation}
where matrix $\bm{A}$ is given by
\begin{equation}
\bm{A}=\frac{h^{3}}{4 \pi}\left[\bm{D H D}^{T}+\operatorname{diag}\{P\}\right]
\end{equation}
with $P$ being the vector of function $p(\bm{r})$, $\bm{G}$ representing the lattice Green's function, $\bm{D}$ being the difference matrices of operators $\nabla$, and $\bm{I}$ being the unit matrix. It can be observed that the solution of the Green's function is equivalent to the matrix inversion, $\bm{G}=\bm{A}^{-1}$. Directly computing the inverse of the matrix is complex and unnecessary. Specially, only diag($\bm{G}$) is needed to obtain the self energy $c({\bm{r}})$. Thus, the SelInvHIF is employed to extract the diagonals of the inverse.

\section{Numerical Results}
\label{Numerical}

To assess the effectiveness of the SelInvHIF, we present numerical results for the MPB equations in three dimensions. The scaling of the computational time is of particular interest. We set the coupling parameter $\Xi=1$ and the uniform fugacity parameter $\Lambda=0.05$. 
For both the PB and the self-consistent iterations, the error criteria are set at $10^{-8}$.
The initial value for the potential in the iteration are always constant in our instances with $\phi^{(0)}=0$. It is important to note that the choice regarding the ID step depends on the problem specified. Both the PB and the GDH stages make use of Dirichlet boundary conditions. The calculation is executed on a machine with Intel Xeon 2.2GHz and 2TB memory. 
All experiments are performed in Matlab with the FLAM package \cite{Ho2020FLAMFL} for hierarchical matrices.
Prior to resolving the MPB equation, we begin by demonstrating an illustrative case of evaluating the diagonal elements of an elliptic differential operator.
The statistical computation time is computed as the average of five measurements. 

\begin{figure}[H]
	\centering
	\includegraphics[width=1\textwidth]{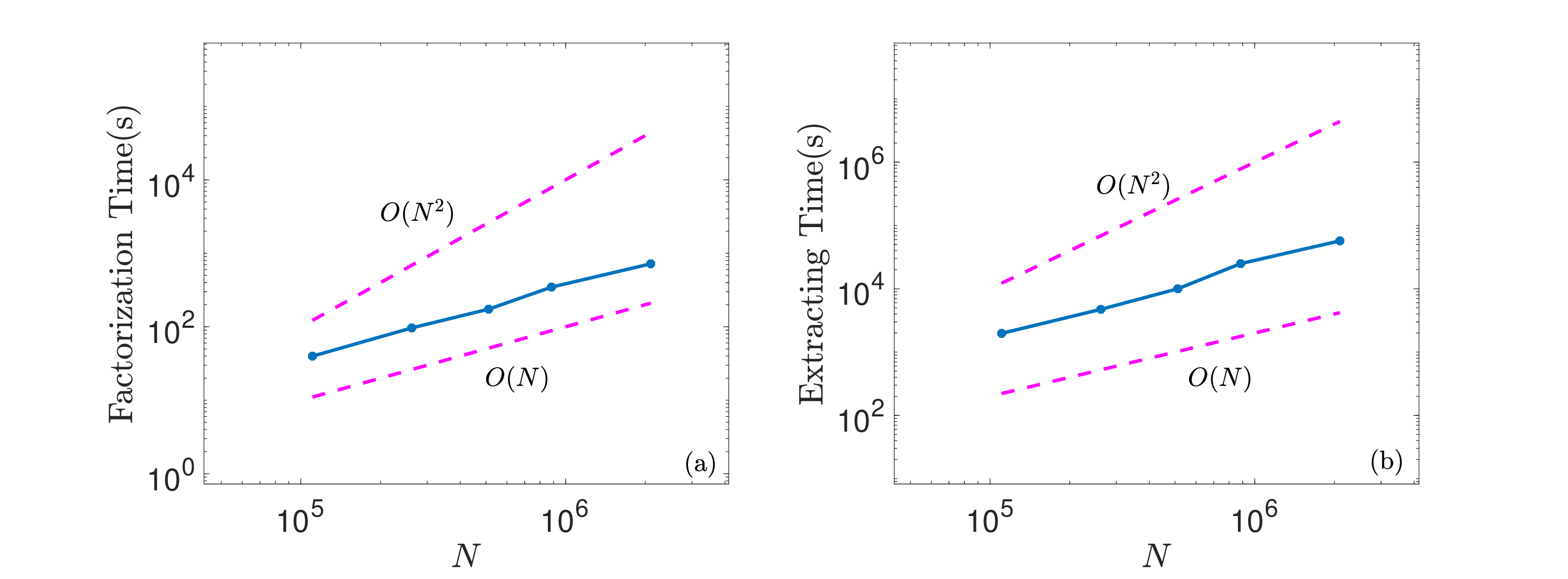}
	\caption{One step time for the SelInvHIF. The solid lines represent the computational time for the factorization step (a) and extracting step (b). The dash lines represents the reference scalings.
}
	\label{scaling}
\end{figure}

{\bf Example 1: The discrete elliptic differential operator (3D).} In our examination of three-dimensional problems, we first explore the diagonals of the inverse of a discrete elliptic differential operator. This is achieved through the implementation of a seven-point stencil discretization. Subsequently, one computes the diagonals of the inverse matrix utilizing both the SelInvHIF method and the exact approach as described in \cite{Lin2009}.
The diagonals of the inverse of the discrete operator are set as $d_s$ and $d_e$, respectively.
Table \ref{Table1} presents the absolute $L^2$ error $E_a=\sqrt{\sum(d_s-d_e)^2/N}$ between the numerical results and the reference solution, which corresponds to the matrix size as determined by the exact method. Additionally, Table \ref{Table1} highlights the relative $L^2$ errors $E_r=\|d_s-d_e\|_2 / \|d_e\|_2$, serving to validate the accuracy of the SelInvHIF. Furthermore, the computational time of the algorithm is displayed in Table \ref{Table1}, while Figure \ref{scaling} confirms the quasi-linear scaling of SelInvHIF.
Table \ref{Table 2} illustrates the relationship between the rank of the ID step and the numerical error. This relationship demonstrates that the error can be effectively controlled as the rank of the ID step increases.  

\begin{table}[H]
\centering
\begin{tabular}{rcccc}
\toprule
\midrule
 Matrix size $\sqrt[3]{N}$ & SelInvHIF time (s)& $E_a$  & $E_r$ \\
\midrule
 \multicolumn{1}{c}{$48$}  & $2.0E + 3$ & $6.5E - 3$& $2.7E - 2$ \\
 \multicolumn{1}{c}{$64$}  & $4.8E + 3$ & $8.1E - 3$  & $3.4E - 2$ \\
 \multicolumn{1}{c}{$80$}  & $1.0E + 4$ &   $9.2E -3$ & $ 3.8E -2$ \\
 \multicolumn{1}{c}{$96$}  & $2.5E + 4$ &   $9.8E - 3$ & $4.0E -2$ \\
 \multicolumn{1}{c}{$128$}  & $5.8E + 4$ &   $ - $ & $ - $ \\
\midrule
\bottomrule
\end{tabular}
\caption{The CPU time, accuracy, and matrix size. The SelInvHIF time means the execution time spent for one step SelInvHIF. The rank in the ID step  takes 37.}
\label{Table1}
\end{table}

\begin{table}[H]
\centering
\begin{tabular}{rcccc}
\toprule
\midrule
The rank of ID step & $E_a$ & $E_r$  \\
\midrule
 \multicolumn{1}{c}{$32$}  &  $3.5E - 3$  & $9.5E - 2$\\
 \multicolumn{1}{c}{$128$}  & $6.5E - 4$ & $8.1E - 3$  \\
 \multicolumn{1}{c}{$256$}  & $7.2E - 8 $ &   $9.2E -7$  \\
 \multicolumn{1}{c}{$512$}  & $5.3E - 16$ &   $9.8E - 15$ \\

\midrule
\bottomrule
\end{tabular}
\caption{Numerical errors for different ranks of the ID step.}
\label{Table 2}
\end{table}

\begin{figure}[H]
	\centering
	\includegraphics[width=0.4\textwidth]{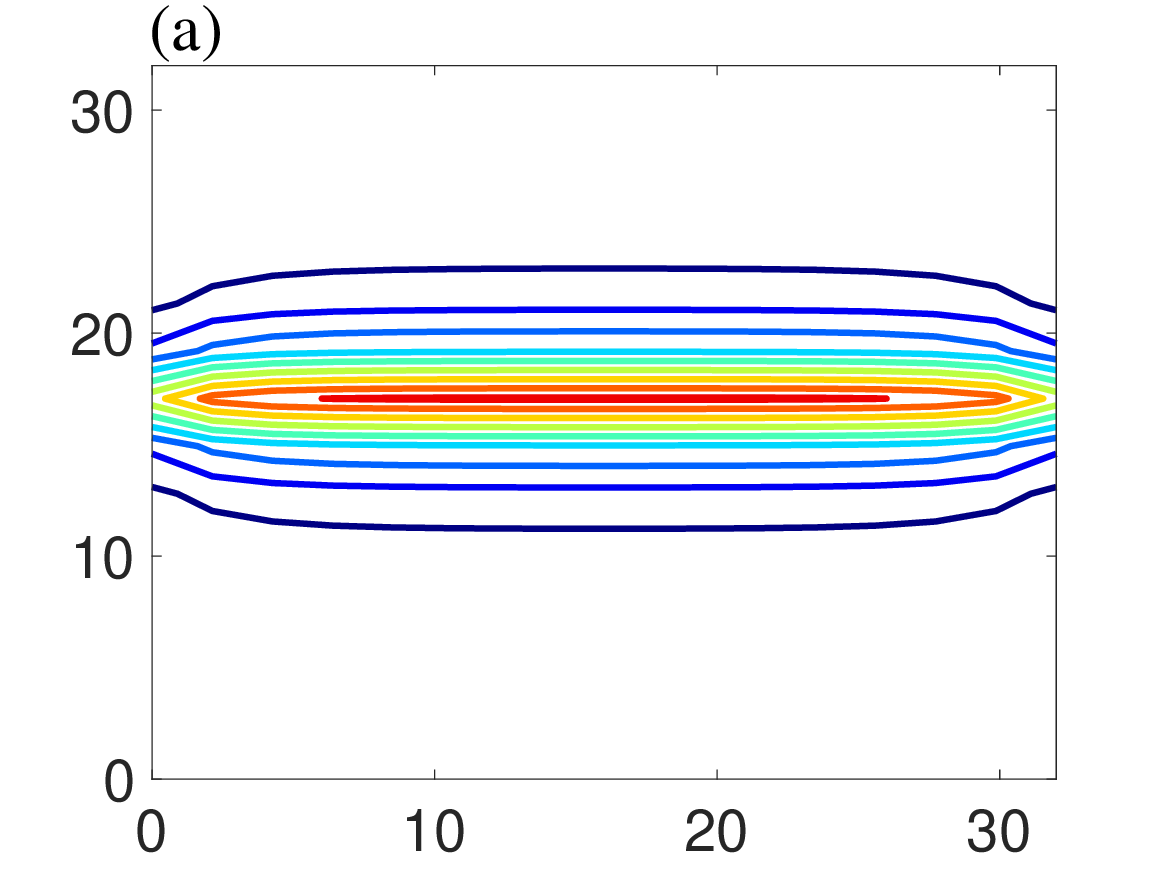}
	\includegraphics[width=0.4\textwidth]{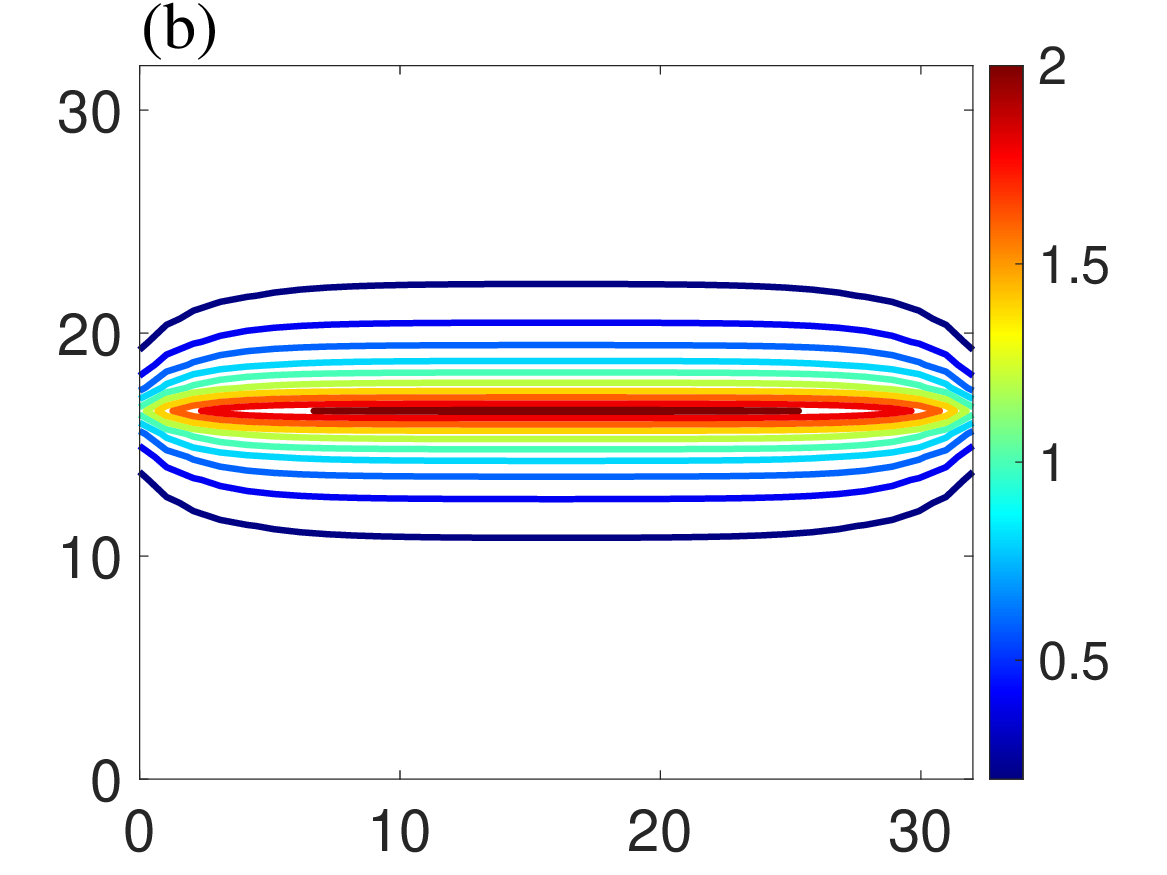}
	\includegraphics[width=0.4\textwidth]{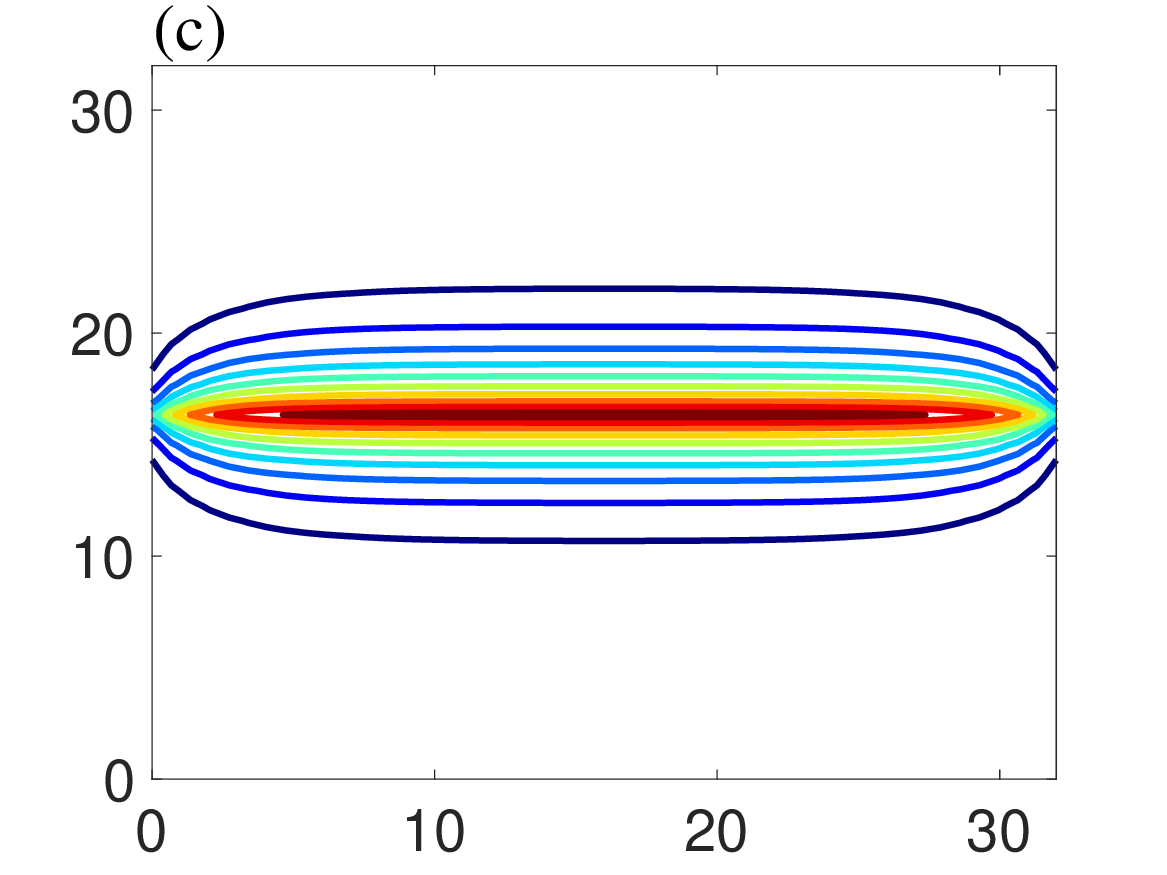}
	\includegraphics[width=0.4\textwidth]{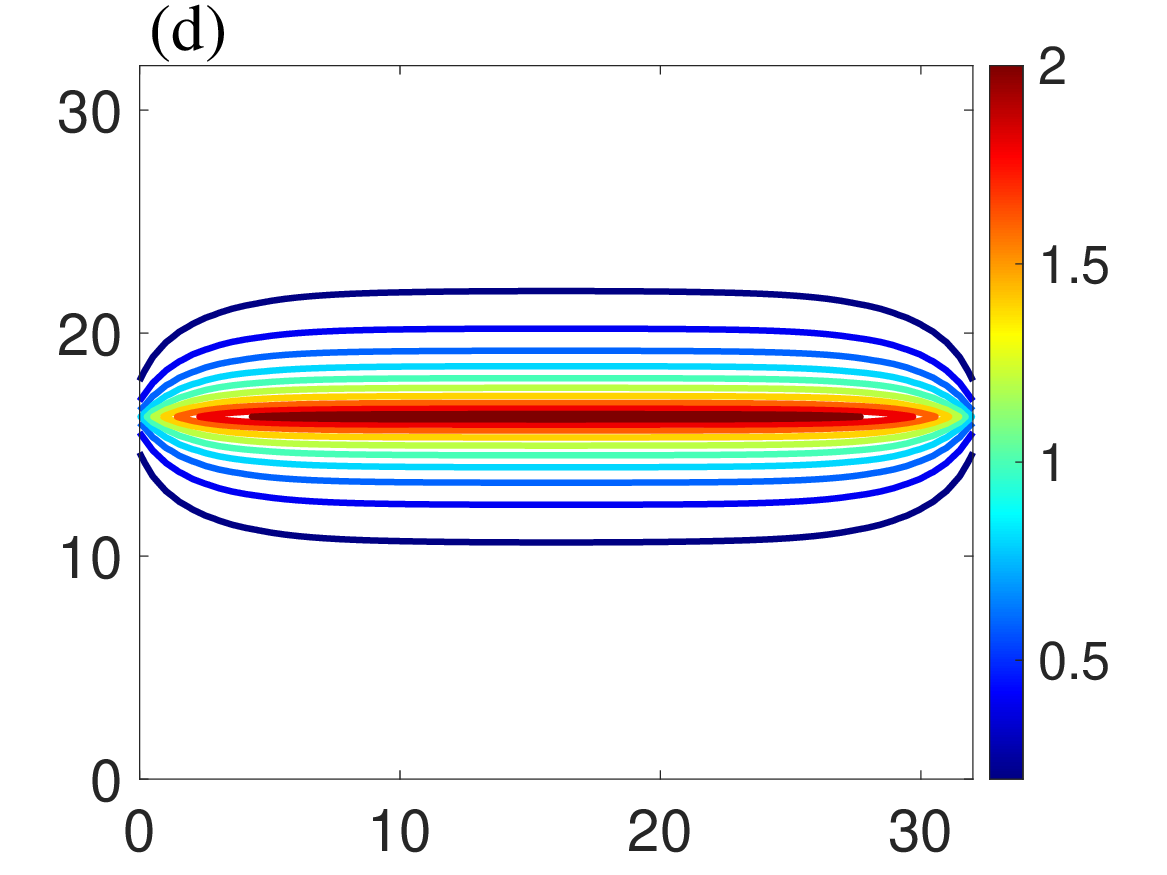}
	\caption{Potential distributions with different matrix size at $z=L/2$.
	 (a): matrix size $N=16$; (b): matrix size $N=32$; (c): matrix size $N=48$; (d): matrix size $N=64$.}
        \label{ex:62}
\end{figure}

{\bf Example 2: The charge density with a delta function.} In this example, we consider discontinuous charged distribution in a region $[0, L]^3$ with $L = 32$. Let the fixed charge density be a face charge:
\begin{equation*}
\rho_{f}(x)=\delta(x-L / 2).
\end{equation*}
One calculates the results of the MPB equations by the SelInvHIF. Figure \ref{ex:62}  visualizes the distribution of the potential in this system at $z=L/2$ with different matrix sizes $N=16^{3}, 32^{3},48^{3}$ and $64^{3}$. 
The potential with respect to $x=L/2$ and $y=L/2$ remains symmetric due to the symmetry of the fixed charge and is most pronounced at $x=L/2$ due to the presence of the charge.
Table \ref{Table 3}  shows the accuracy of the whole algorithm to compute the potential $\phi$ compared to a reference potential computed with a sufficiently large grid size $N=64^3$, which verifies the approximate accuracy of first-order due to the discontinuous of the derivative of the potential at $y=z=L/2$. Furthermore, Table \ref{Table 3} also shows the computational time of the algorithm to verify the quasilinear scaling of the SelInvHIF.

\begin{table}[h]
\centering
\begin{tabular}{rcccc}
\toprule
\midrule
Matrix size $\sqrt[3]{N}$ & Total time & SelInvHIF time & $L_{\infty}$ error\\
\midrule
 \multicolumn{1}{c}{$16$}  & $4.02E +  0$ & $2.83E + 0$ & $5.56E - 1$  \\
  \multicolumn{1}{c}{$32$}   & $2.12E + 2$ & $1.89E + 2$ & $3.23E - 1$   \\
  \multicolumn{1}{c}{$48$}   & $2.08E + 3$ & $1.95E + 3$ & $1.33E - 1$ \\
  \multicolumn{1}{c}{$64$}   & $7.74E + 3$ & $6.92E + 3$ & - \\
\midrule
\bottomrule
\end{tabular}
\caption{The CPU time, accuracy, and matrix size. The total time and the SelInvHIF time mean the execution time spent for one step iteration in the whole program and the time for one step SelInvHIF, respectively.}
\label{Table 3}
\end{table}

%
%
%

\begin{figure}[H]
	\centering
	\includegraphics[width=0.4\textwidth]{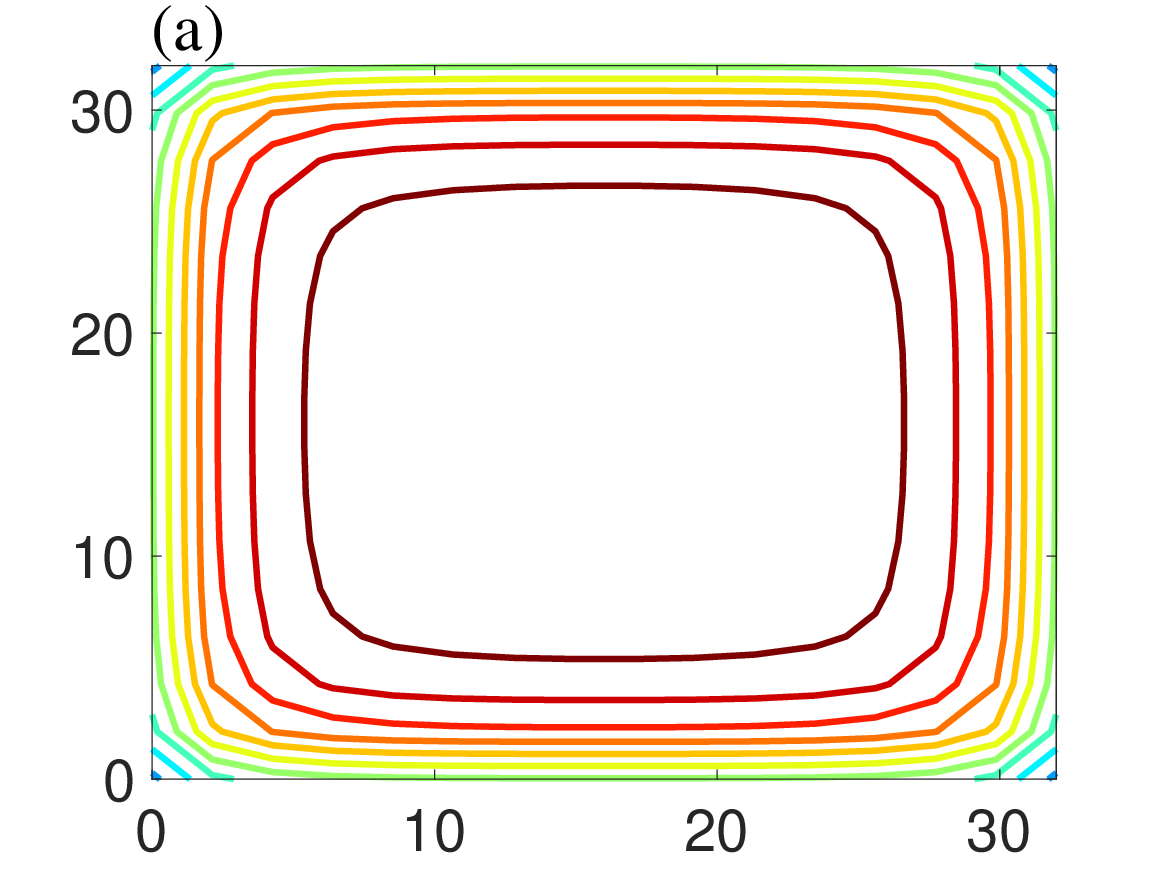}
	\includegraphics[width=0.4\textwidth]{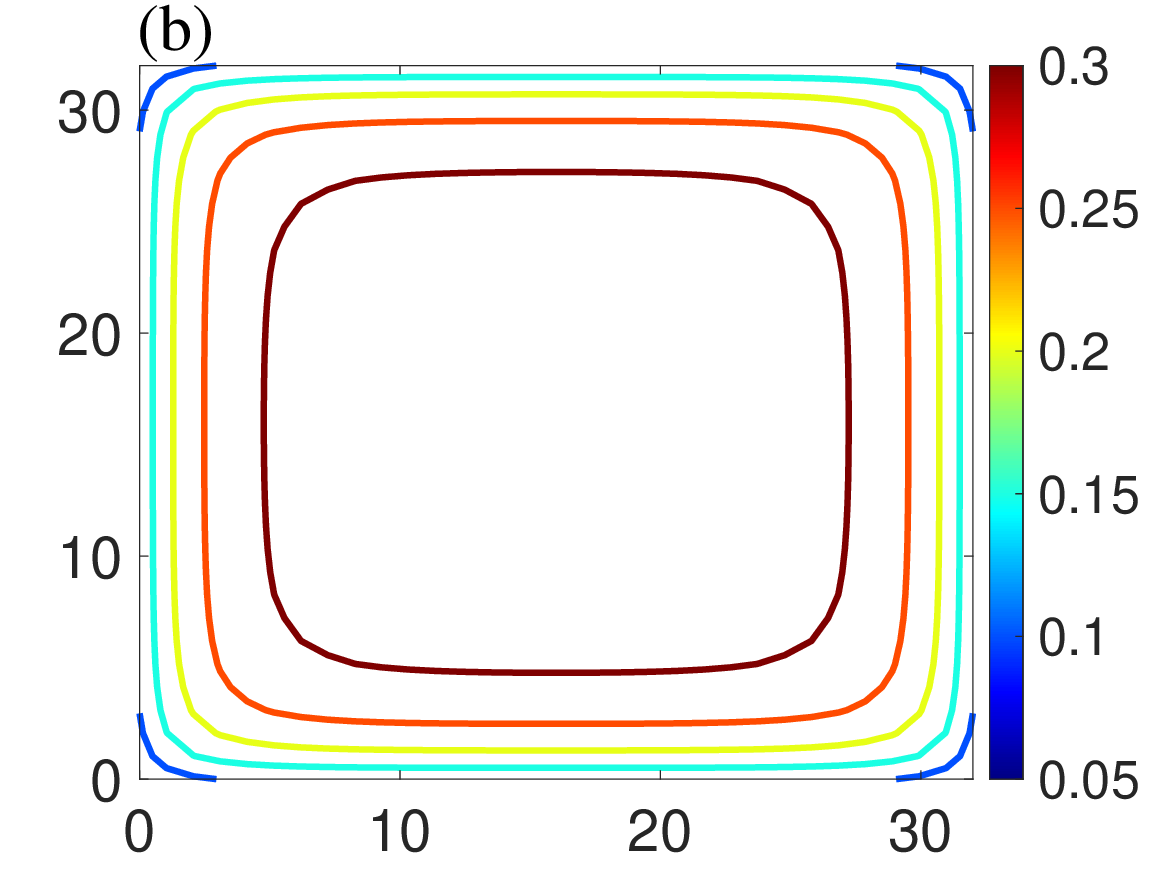}
	\includegraphics[width=0.4\textwidth]{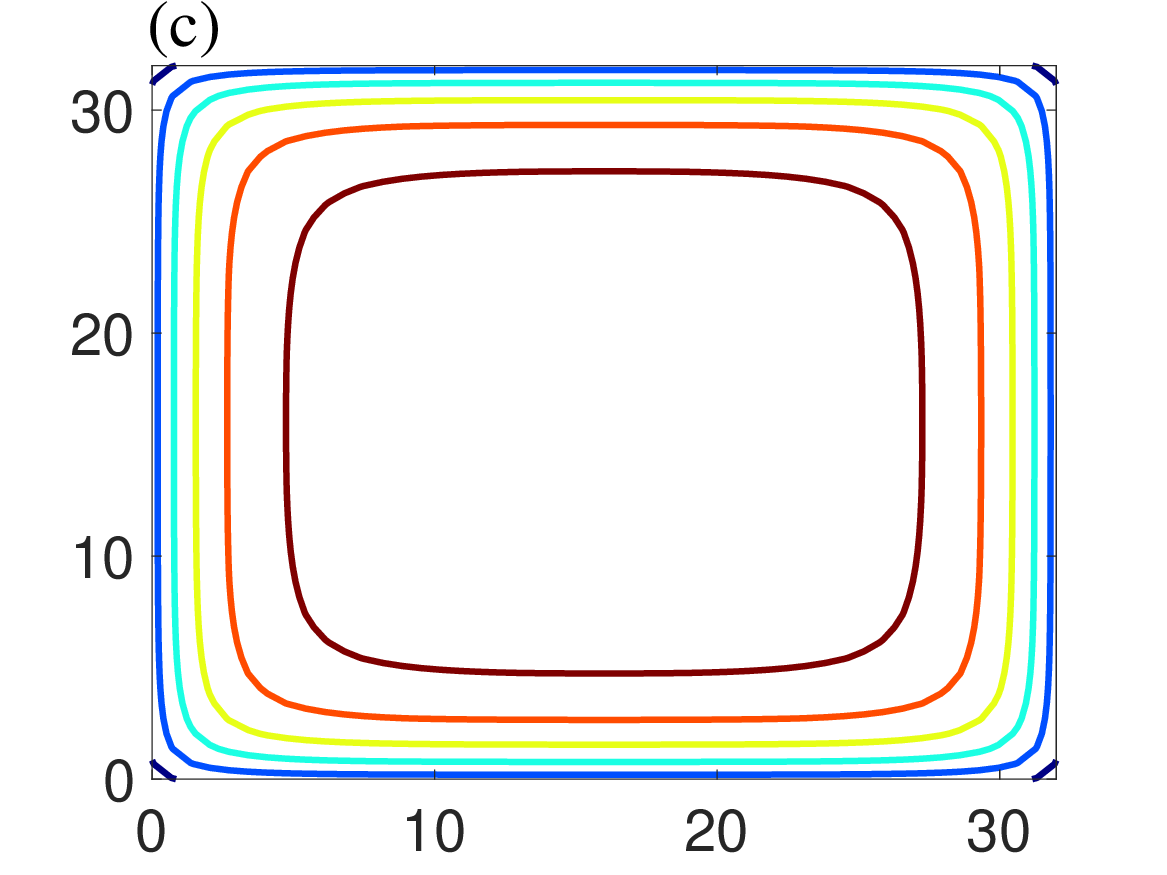}
	\includegraphics[width=0.4\textwidth]{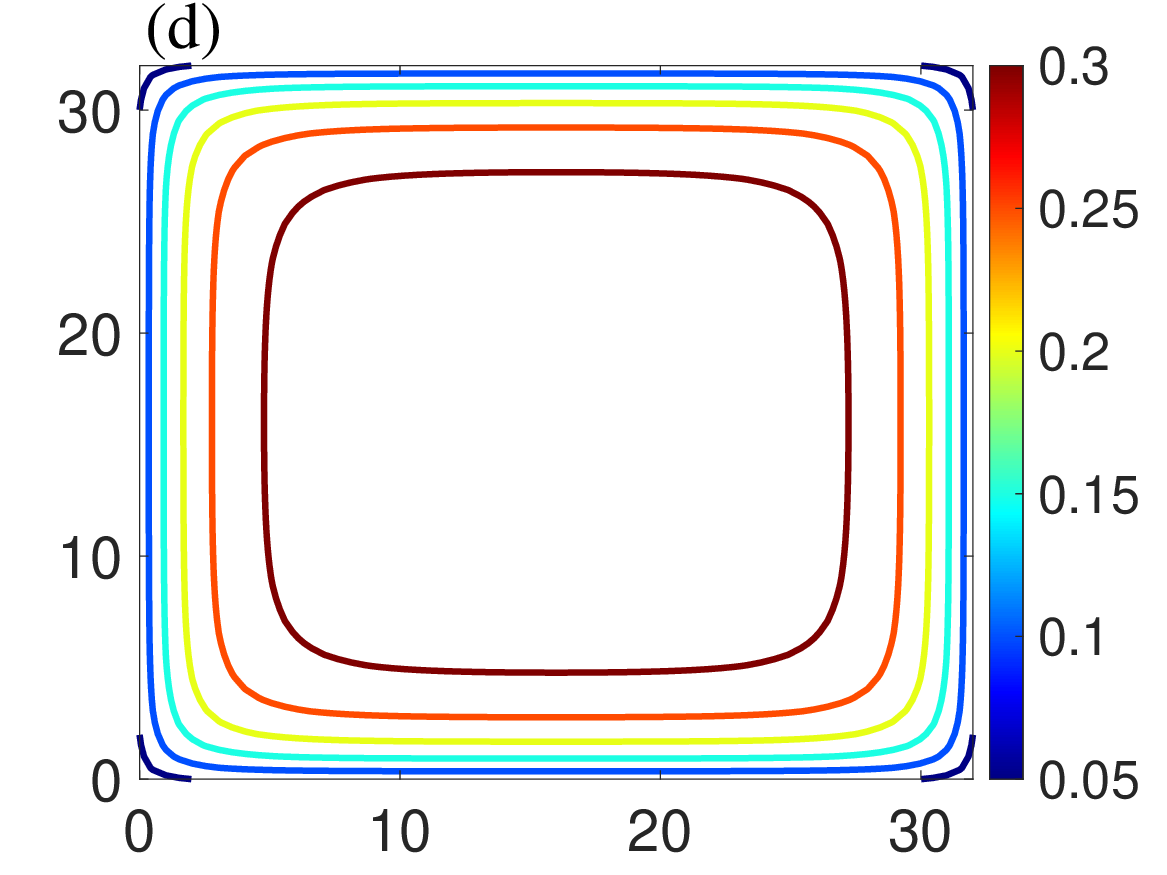}
	\caption{Potential distributions with different matrix size at $z=L/2$.
	 (a): matrix size $N=16$; (b): matrix size $N=32$; (c): matrix size $N=48$; (d): matrix size $N=64$.}
        \label{ex:3}
\end{figure}

{\bf Example 3: The charge density with continuous function.} In this example, we consider discontinuous charged distribution in a region $[0, L]^3$ with $L = 32$. Let the fixed charges density be:
\begin{equation*}
\rho_{f}(x)=sin(\pi x/L).
\end{equation*}
One then calculates the results of the MPB equations by SelInvHIF with the accuracy $10^{-8}$ in the ID step. 
Figure \ref{ex:3}  visualizes the distribution of the potential in this system at $x=L/2$ with different matrix sizes $N=16^{3}, 32^{3},48^{3}$ and $64^{3}$, which displays the convergence.
The potential are most evident at $x = L/2$ due to the dense charge.
Table \ref{Table 4}  shows the accuracy of the whole algorithm to compute the potential $\phi$ compared to a reference potential computed with a sufficiently large grid size $N=64^3$, which verifies the convergence of our algorithm. Furthermore, Table \ref{Table 4} also shows the computational time of the algorithm to verify the quasilinear scaling of the SelInvHIF. 


\begin{table}[H]
\centering
\begin{tabular}{rcccc}
\toprule
\midrule
Matrix size $\sqrt[3]{N}$ & Total time & SelInvHIF time & $L_{\infty}$\ \\
\midrule
 \multicolumn{1}{c}{$16$}  &  $2.75E+1$  & $2.35E+1$  & $2.12E-2$  \\
 \multicolumn{1}{c}{$32$}  & $3.06E +2$    &  $2.47E+2$  & $3.84E-3$ \\
 \multicolumn{1}{c}{$48$}  &  $2.80E+3$     & $2.40E+3$  & $1.92E-3$   \\
  \multicolumn{1}{c}{$64$}   &  $1.06E+4$   & $8.37E+3$ & $-$    \\

\midrule
\bottomrule
\end{tabular}
\caption{The CPU time, accuracy, and matrix size. The total time and the SelInvHIF time mean the execution time spent for one step iteration in the whole program and the time for one step SelInvHIF, respectively.}
\label{Table 4}
\end{table}

\section{\large Conclusions}
\label{Conclusion}
This paper develops the SelInvHIF , a fast algorithm to solve the MPB equations.
The SelInvHIF effectively integrates hierarchical interpolative factorization and selected inverse techniques to achieve an $O(N\log N)$ computational complexity and $O(N)$ computational complexity with edge skeletonization, in terms of operations and memory, necessary for computing the diagonal of the inverse of a sparse matrix discretized from an elliptic differential operator. The proposed algorithm was applied to three-dimensional MPB problems, and demonstrated impressive performance in terms of both accuracy and efficiency. 

\section*{Acknowledgment}
Y. Tu and Z. Xu acknowledge the financial support from the National Natural Science Foundation of China (grant No. 12071288), Science and Technology Commission of Shanghai Municipality (grant Nos. 20JC1414100 and 21JC1403700) and Strategic Priority Research Program of Chinese Academy of Sciences (grant No. XDA25010403). H. Yang thanks the support of the US National Science Foundation under award DMS-1945029.
\bibliographystyle{unsrt}
\bibliography{ref}

\end{document}